 \documentclass[final,3p,times,twocolumn]{elsarticle}

\usepackage{amssymb}

\bibpunct{(}{)}{,}{a}{}{,}

\journal{Icarus}

\usepackage{endfloat}
\usepackage{ulem}

  \newcommand{\ie}{i.e.}
  \newcommand{\eg}{e.g.}
  \newcommand{\gcm}{g/cm$^{3}$}
  \newcommand{\degr}{\ensuremath{^\circ}}
  \newcommand{\astar}{\ensuremath{a^\star}}

  \newcommand{\add}[1]{#1}

\usepackage{hyperref}

\graphicspath{{figures/}}

\begin{document}

\begin{frontmatter}

\title{The taxonomic distribution of asteroids from multi-filter all-sky photometric surveys}

\author[mit]{F. E. DeMeo}
\ead{fdemeo@mit.edu}
\author[imcce,esac]{B. Carry}

\address[mit]{Department of Earth, Atmospheric and Planetary Sciences, MIT, 77 Massachusetts Avenue, Cambridge, MA, 02139, USA}
\address[imcce]{Institut de M{\'e}canique C{\'e}leste et de Calcul des {\'E}ph{\'e}m{\'e}rides, Observatoire de Paris, UMR8028 CNRS, 77 av. Denfert-Rochereau 75014 Paris, France}
\address[esac]{European Space Astronomy Centre, ESA, P.O. Box 78, 28691 Villanueva de la Ca{\~n}ada, Madrid, Spain}

\begin{abstract}
  The distribution of asteroids across the Main Belt has been studied for 
  decades to understand the current compositional distribution and what 
  that tells us about the formation and evolution of our solar system. 
  All-sky surveys now provide orders of magnitude more data than targeted 
  surveys. We present a method to bias-correct the asteroid population
  observed in the Sloan Digital Sky Survey (SDSS) according
  to size, distance, and albedo. We taxonomically classify this dataset consistent
  with the Bus  \citep{2002-Icarus-158-BusII}
  and Bus-DeMeo \citep{2009-Icarus-202-DeMeo} systems and 
  present the resulting taxonomic distribution. The dataset includes asteroids 
  as small as 5\,km, a factor of three in diameter smaller than in previous work
  such as by \citet{2003-Icarus-162-Mothe-Diniz}. 
  \add{Because of the wide range of sizes in our sample, we present the distribution 
    by number, surface area, volume, and mass whereas previous work was exclusively by number. While 
    the distribution by number is a useful quantity and has been used for decades, these 
    additional quantities provide new insights into the
    distribution of total material.} 
  \add{We find evidence for D-types in the inner main belt where they are
    unexpected according to dynamical models of implantation of bodies from
    the outer solar system into the inner solar system during planetary migration
    \citep{2009-Nature-460-Levison}.
    We find no evidence of S-types or other unexpected classes among Trojans and Hildas, albeit a
    bias favoring such a detection.
    Finally, we estimate
    for the first time
    the total amount of material of each class in the inner
    solar system. The main belt's most massive classes are C, B, P, V and S in 
    decreasing order. Excluding the four most massive asteroids, (1) Ceres, (2) Pallas, (4) Vesta and
    (10) Hygiea that heavily skew the values, 
    primitive material (C-, P-types) account for more than half main-belt and
    Trojan asteroids by mass, most of the remaining mass being in the S-types. All the
    other classes are minor contributors to the material between Mars and Jupiter.}
\end{abstract}

\begin{keyword}
Asteroids, surfaces \sep Asteroids, composition \sep spectrophotometry


\end{keyword}

\end{frontmatter}

\newpage

\section{Introduction}


  \indent The current compositional makeup and distribution of bodies in the
  asteroid belt is both a remnant of our early solar system's primordial
  composition and temperature gradient and its subsequent physical and dynamical
  evolution. The distribution of material of different compositions has been studied based on 
  photometric color and spectroscopic studies of $\sim$2,000
  bodies in visible and near-infrared wavelengths
  \citep{1971-NASSP-Chapman, 1975-Icarus-25-Chapman, 1982-Science-216-Gradie,1989-AsteroidsII-Gradie,
    1999-PhD-Bus, 
    2002-Icarus-158-BusII, 
    2003-Icarus-162-Mothe-Diniz}.
  \add{These data were based on all available spectral
    data at the time the work was performed including spectral surveys such as 
    \citet{1984-PhD-Tholen},
    \citet{1985-Icarus-61-Zellner},
    \citet{1987-Icarus-72-Barucci},
    \citet{1995-Icarus-115-Xu},
    \citet{2002-Icarus-158-BusII},
    and \citet{2004-Icarus-172-Lazzaro}.}\\
  \indent \add{The first in-depth study showing the significance of global trends across 
    the belt looked at surface reflectivity (albedo) and spectrometric measurements 
    of 110 asteroids. It was then that the dominant trend in the belt was found: S-types 
    are more abundant in the part of the belt closer to the sun and the C-types further 
    out \citep{1975-Icarus-25-Chapman}. Later work by \citet{1982-Science-216-Gradie} 
    and \citet{1989-AsteroidsII-Gradie} revealed clear trends for each of the major classes 
    of asteroids, concluding that each group formed close to its current
    location.} \\
  \indent \add{The Small Main-belt Asteroid Spectroscopic Survey
    \citep[SMASSII, ][]{2002-Icarus-158-BusI}  
    measured visible spectra for 1,447 asteroids and the Small Solar
    System Objects Spectroscopic 	 
    Survey (S3OS2) observed 820 asteroids \citep{2004-Icarus-172-Lazzaro}.
    The conclusion
    of these major spectral surveys brought new discoveries and views of the main belt. 	 
    \citet{2002-Icarus-158-BusI} found the distribution to be largely
    consistent with \citet{1982-Science-216-Gradie},	  
    however they noted more finer detail within the S and C complex distributions, 
    particularly a secondary peak for C-types at 2.6 AU and for S-types at 2.85 AU. 	 
    \citet{2003-Icarus-162-Mothe-Diniz} combined data from multiple spectral surveys 
    looking at over 2,000 asteroids with H magnitudes smaller than 13 (D$\sim$15\,km 	 
    for the lowest albedo objects). Their work differed from early surveys finding that 	 
    S-types continued to be abundant at further distances, particularly at the smaller 	 
    size range covered in their work rather than the steep dropoff other
    surveys noted. } \\
  \indent Only in the past decade have large surveys at visible and
  mid-infrared wavelengths been available allowing us to tap into the
  compositional detail of the million or so asteroids greater than 1 kilometer
  that are expected to exist in the belt
  \citep{2005-Icarus-175-Bottke}.
  The results of these surveys
  (including discovery surveys), however, are heavily biased toward the closest,
  largest, and brightest of asteroids. This distorts our overall picture of the belt
  and affects subsequent interpretation.\\
  \indent In this work we focus on the data from
  the Sloan Digital Sky Survey Moving Object Catalog
  \citep[SDSS, MOC,][]{2001-AJ-122-Ivezic,2002-AJ-124-Ivezic} that observed over 
  100,000 unique asteroids in five photometric bands over visible wavelengths.
  These bands provide enough information to broadly classify these objects
  taxonomically \citep[\eg,][]{2010-AA-510-Carvano}.
  \add{In this work we refer to the SDSS MOC as SDSS for simplicity.} 
  We classify the SDSS data and determine 
  the distribution of asteroids in the main belt.  We present a method to correct for the 
  survey's bias against the dimmest, furthest bodies.\\
  \indent Traditionally, the asteroid compositional distribution has
  been shown as the number objects of each taxonomic type as function of
  distance. 
  \add{While the number distribution is important for size-frequency distributions
  and understanding the collisional environment in the asteroid belt}, 
  the concern with this method is that objects of very different sizes are weighted equally. 
  For example, objects with diameters ranging from 15\,km to greater than 500\,km 
  were assigned equal importance in previous works. This is particularly
  troublesome for SDSS and other large surveys because the distribution by number
  further misrepresents the amount of material of each class by equally 
  weighting objects that differ by \add{two orders of magnitude in diameter} and 
  by six orders of magnitude in volume.
  To create a more realistic and comprehensive view of the asteroid belt  
  we provide the taxonomic distribution according to number, surface area,
  volume, and mass.
  \add{New challenges are presented when attempting to create
  these distributions including
  the inability to account for the smallest objects (below the efficiency limit of SDSS),
  the incompleteness of SDSS even at size ranges where the survey is efficient, and 
  incomplete knowledge of the exact diameters, albedos and densities of each
  object. We attempt to correct for as many of these issues as possible in the
  present study.}\\
  \indent The distribution according to surface area 
  is perhaps the most technically correct result because only the surfaces of
  these bodies are measured.  
  \add{We only have indirect information about asteroid interiors, mainly derived from
    the comparison of their bulk density with that of their surface material,
    suggesting differentiation in some cases, and presence of voids in others
    \citep[\eg,][]{2008-ChEG-68-Consolmagno, 2012-PSS-73-Carry}.
    The homogeneity in surface reflectance and albedo of
    asteroids pertaining to dynamical families 
    \citep[\eg,][]{2002-AJ-124-Ivezic, 2002-AsteroidsIII-5.1-Cellino,
      2008-Icarus-198-Parker, 2013-MNRAS-Carruba} however suggest that most 
    asteroids have an interior composition similar to their surface
    composition. }
  \add{Nevertheless, recent models find that large bodies even though
    masked with fairly primitive surfaces could actually have differentiated
    interiors \citep{2011-EPSL-305-Elkins-Tanton,2012-PSS-66-Weiss}. }
  The distribution of surface area is 
  relevant for dust creation from non-catastrophic collisions
  \add{\citep[e.g.][]{2006-Icarus-181-Nesvorny,2008-ApJ-679-Nesvorny}}
  and from a resource standpoint such 
  as for mining materials on asteroid surfaces. The volume of material provides context 
  for the total amount of material in the asteroid belt with surfaces of a
  given taxonomic class. While we  
  do not know the actual composition or properties of the interiors we can at least account 
  for the material that exists. \\
  \indent The most ideal case is to determine the distribution of \textsl{mass}.
  This view accounts for all of the material in the belt, corrects for composition and porosity 
  of the interior and properly weights the relative importance of each asteroid according
  to size and density. While the field is a long way away from having perfectly detailed
  shape and density measurements for every asteroid, by applying estimated sizes and 
  average densities per taxonomic class to a large, statistical sample, we provide 
  in this work the first look at the distribution of classes in the asteroid
  belt according to mass\add{, and estimate the total
    amount of material each class represents in the inner solar system.} \\
  \indent \add{The next section (Sec~\ref{sec: data}) introduces the data used for this work.}
  We overview observing biases and our correction method in
  Sections~\ref{sec: bias} and~\ref{sec: subset}.
  We describe our classification method for our sample in Section~\ref{sec: taxo}. 
  We then explain in Section~\ref{sec: distribuild} our method  
  for building the compositional distribution and application of our dataset
  to all asteroids in the main belt.
  Finally, we present in Section~\ref{sec: distrib}
  the bias-corrected taxonomic distribution of   
  asteroid material across the main belt according to number, surface area,
  volume, and mass\add{, and discuss the results in
    Section~\ref{sec: discuss}.}

\section{The Dataset\label{sec: data}}

  \subsection{Selection of high quality measurements from SDSS\label{sec: selectSDSS}}

\begin{figure*}[t]
  \includegraphics[width=\textwidth]{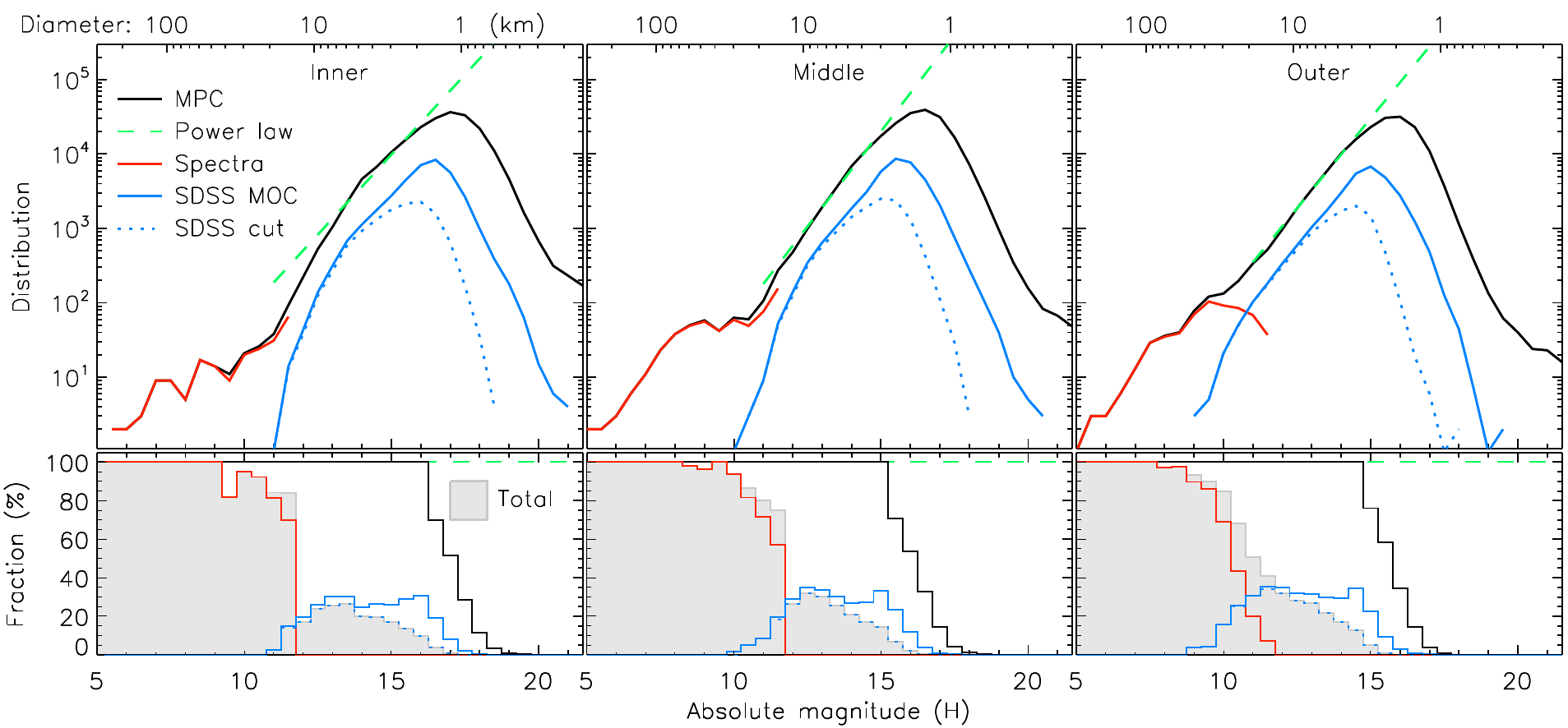}
  \caption[Completeness of SDSS measurements]{%
    Number (top) and fraction (bottom) of all asteroids discovered (solid
    black) and observed by spectroscopic surveys (red), or SDSS MOC (blue) in
    each zone of the main belt.
    The subset of SDSS MOC we used here
    (with cuts applied to photometry, see Section~\ref{sec: selectSDSS})
    is shown in dotted blue. The 
    completeness of discovered asteroids at each size range is determined by
    extrapolating the expected population using a power law fit (dashed green)
    to the MPC list of discovered asteroids (solid black). The power law indices
    \add{calculated in this work (see Sec.~\ref{sec: SDSStoMPC})}
    for the IMB, MMB, and OMB (determined over the H magnitude range
    14-16, 13-15, and 12-14.5) are -2.15, -2.57, and -2.42, respectively. In
    the bottom panel the total fraction of the sample (before bias correction)
    is shaded in gray. 
    \label{fig: sdsscomplete}
  }
\end{figure*}

\begin{figure*}[t]
  \includegraphics[width=\textwidth]{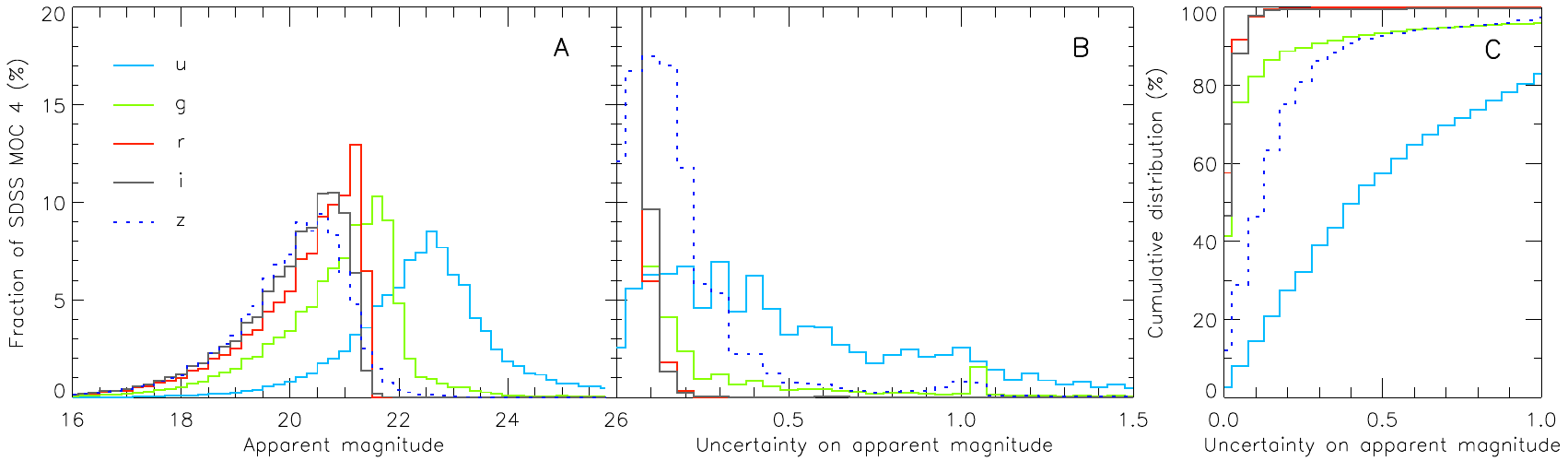}
  \caption[Uncertainty on SDSS photometry]{%
    Distribution of the apparent magnitude (A) and associated uncertainty (B)
    for all the SDSS MOC4 observations (471,569). The larger uncertainty
    affecting the observations in the u' filter (C) precludes any reliable
    classification information to be retrieved from this filter. 
    \label{fig: filters}
  }
\end{figure*}

  \indent The Sloan Digital Sky Survey (SDSS) is an imaging \add{and
    spectroscopy} survey dedicated to 
  observing galaxies and quasars \citep{2001-AJ-122-Ivezic}.
  The images are taken in 5 filters, u', g', r', i', and z', from 0.3 to
  1.0\,$\mu$m. 
  The survey also observed over 400,000 moving objects in our solar system of which over
  100,000 are unique objects linked to known asteroids. The current release
  of the Moving Object Catalogue \citep[SDSS MOC 4,][]{2002-AJ-124-Ivezic}
  includes observations through March 2007.\\
  \indent We restrict our sample from the SDSS MOC database according to the
  following criteria. First, we keep only objects assigned a number or a
  provisional designation, \ie, those for which we can retrieve the orbital
  elements.
  We then remove observations that are deemed unreliable: with any
  apparent magnitudes greater than 22.0, 22.2, 22.2, 21.3, 20.5 for each
  filter (5.9\% of the SDSS MOC4),
  which are the limiting magnitudes for 95\% completeness 
  \citep{2001-AJ-122-Ivezic},
  or any photometric uncertainty greater than 0.05 (excluding the u' filter, explained below). These
  constraints remove a very large portion of the SDSS dataset (about 87\% of
  all observations), largely due to the greater typical error for the
  z' filter. While there is only a small subset of the sample remaining   
  (Fig.~\ref{fig: sdsscomplete}),
  we are assured of the quality of the data. Additionally,
  for higher errors, the ambiguity among taxonomic classes possible for an
  object becomes so great that any classification becomes essentially
  meaningless. We exclude the u' filter from this work primarily
  because of the significantly higher errors in this filter compared to the
  others (Fig.~\ref{fig: filters}),
  and secondarily because neither the Bus nor Bus-DeMeo taxonomies
  \citep[that we use as reference for classification
    consistency,][]{2002-Icarus-158-BusII,2009-Icarus-202-DeMeo} 
  covered that wavelength range.  
    
    The fourth release of the MOC contains non-photometric nights in the
    dataset. The SDSS provides data checks that indicate potential problems
    with the measurements\footnote{%
      \href{http://www.sdss.org/dr4/products/catalogs/flags.html}%
           {http://www.sdss.org/dr4/products/catalogs/flags.html},
      \href{http://www.sdss.org/dr4/products/catalogs/flags_detail.html}%
           {http://www.sdss.org/dr4/products/catalogs/flags\_detail.html},
      \href{http://www.sdss.org/dr7/tutorials/flags/index.html}%
           {http://www.sdss.org/dr7/tutorials/flags/index.html},
      \href{http://www.astro.washington.edu/users/ivezic/sdssmoc/moving_flags.txt}%
           {http://www.astro.washington.edu/users/ivezic/sdssmoc/moving\_flags.txt}.
    },
    and we thus remove observations with flags relevant
    to moving objects and good photometry: 
    \texttt{edge},
    \texttt{badsky},
    \texttt{peaks too close},
    \texttt{not checked},
    \texttt{binned4}, 
    \texttt{nodeblend},
    \texttt{deblend degenerate}, 
    \texttt{bad moving fit}, 
    \texttt{too few good detections}, and
    \texttt{stationary}.
    \add{These flags note issues such as data where
    objects were too close
    to the edge of the frame, the peaks from two objects were too close to be deblended,
    the object was detected only in a 4x4 binned frame, or the object was not detected
    as moving. }
    \add{Further details of the flags are provided on the websites in the footnotes.}
    \add{The presence of these flags does not necessary imply problematic
      data, but because the observations removed due to these
    flags represent a small percentage of the total objects that fall within
    the magnitude and photometric error constraints ($\sim$2\%), we
      prefer to slightly restrict the sample than to contaminate it}.
    Of the 471,569
    observations in MOC4 we have a sample of 58,607 observations after
    applying the selection criteria. We keep observations that are flagged as
    having interpolation (37\% of our sample), including
    \texttt{psf flux interp} (26\% of our sample) which indicates that over
    20\% of the point spread function 
    flux is interpolated. We also include observations corrected for cosmic
    rays (6.5\%) and those that might have a cosmic ray but are uncorrected 
    (1.5\%).
    Anyone wishing to use the SDSS data or classification results to analyze particular
    objects rather than large populations is cautioned to note all flags
    associated with an observation.

  \subsection{Average albedo of each taxonomic class\label{sec: albedo}}
    \indent There have been recent efforts to determine average albedos per
    taxonomic class
    \citep{2010-AJ-140-Ryan, 2011-PASJ-63-Usui, 2011-ApJ-741-Masiero}. These 
    results can be used to more accurately estimate the diameter of a body
    of a given taxonomic class.
    In some cases, however, the results disagree by more
    than the reported uncertainties (\eg, B-types, see
    Table~\ref{tab: albedo}). \add{We
    calculate mean values,
    weighted by the number of albedos determined and their accuracy, 
    for each taxonomic class for this
    work based on averages reported from previously published results
    \citep{2010-AJ-140-Ryan, 2011-PASJ-63-Usui, 2011-ApJ-741-Masiero}.}
    See Table~\ref{tab: albedo} for a summary of published values and the 
    averages we use in this work. It must also be noted that the average albedo per class
    does not necessarily represent the actual albedo for any particular object
    because albedo may vary greatly among each class
    \citep[\eg,][]{2011-ApJ-741-Masiero}. 
    
    The X class is divided into three classes, E, M, and P,
    distinguished solely by their albedo
    (P\,$<$\,0.075, 0.075\,$<$\,M\,$<$\,0.30, E\,$>$\,0.30). We calculate
    average albedo in each 
    class from the roughly 2,000 objects in our sample that have WISE, AKARI,
    or IRAS albedos. We find average albedos of
    0.45, 0.14, and 0.05
    for E, M, and P, respectively. Because the average albedo for a given class is
    calculated solely using objects with spectral data, and the spectral
    measurements are biased toward brighter, higher albedo objects, this
    average could consequently be biased toward higher albedos.

\begin{table*}
\begin{center}
\begin{tabular}{c|rcc|rc|rc|c}
\hline
\hline
 & \multicolumn{3}{c|}{IRAS}
     & \multicolumn{2}{c|}{AKARI}
     & \multicolumn{2}{c|}{WISE}
     & \multicolumn{1}{c}{Average} \\
 Class & \# & a$_{\textrm{\small STM}}$ & a$_{\textrm{\small NEATM}}$ &
         \# & a &
         \# & a & a \\
\hline
  A    &  4 & 0.26\,$\pm$\,0.12 & 0.18\,$\pm$\,0.04 &   6 & 0.23\,$\pm$\,0.06 &   5 & 0.19\,$\pm$\,0.03 & 0.20\,$\pm$\,0.03 \\
  B    &  2 & 0.26\,$\pm$\,0.13 & 0.08\,$\pm$\,0.09 &   3 & 0.14\,$\pm$\,0.03 &   2 & 0.12\,$\pm$\,0.02 & 0.14\,$\pm$\,0.04 \\
  C    & 42 & 0.08\,$\pm$\,0.02 & 0.06\,$\pm$\,0.01 &  44 & 0.06\,$\pm$\,0.03 &  32 & 0.06\,$\pm$\,0.03 & 0.06\,$\pm$\,0.01 \\
  D    & 11 & 0.08\,$\pm$\,0.03 & 0.07\,$\pm$\,0.03 &  14 & 0.06\,$\pm$\,0.03 &  13 & 0.05\,$\pm$\,0.03 & 0.06\,$\pm$\,0.01 \\
  K    & 12 & 0.16\,$\pm$\,0.07 & 0.12\,$\pm$\,0.04 &  14 & 0.14\,$\pm$\,0.04 &  11 & 0.13\,$\pm$\,0.06 & 0.14\,$\pm$\,0.02 \\
  L    & 12 & 0.14\,$\pm$\,0.04 & 0.11\,$\pm$\,0.04 &  16 & 0.12\,$\pm$\,0.04 &  19 & 0.15\,$\pm$\,0.07 & 0.13\,$\pm$\,0.01 \\
  Q    &  1 & 0.51\,$\pm$\,0.10 & 0.41\,$\pm$\,0.08 &   1 & 0.28\,$\pm$\,0.01 &   1 & 0.15\,$\pm$\,0.03 & 0.27\,$\pm$\,0.08 \\
  S    & 50 & 0.26\,$\pm$\,0.06 & 0.20\,$\pm$\,0.06 & 104 & 0.23\,$\pm$\,0.05 & 121 & 0.22\,$\pm$\,0.07 & 0.23\,$\pm$\,0.02 \\
  V    &  1 & 0.37\,$\pm$\,0.08 & 0.35\,$\pm$\,0.07 &   1 & 0.34\,$\pm$\,0.01 &   8 & 0.36\,$\pm$\,0.10 & 0.35\,$\pm$\,0.01 \\
\hline 
\end{tabular}
  \caption[Summary of albedo determinations]{%
    Summary of albedo determination for asteroids listed in
    \citet{2009-Icarus-202-DeMeo} based on 
    radiometry, using data from
    IRAS \citep{2010-AJ-140-Ryan}, 
    AKARI \citep{2011-PASJ-63-Usui}, and
    WISE \citep{2011-ApJ-741-Masiero}
    infrared satellites. The two columns for IRAS correspond to two different
    thermal models applied to the data set
    \citep[STM and NEATM, see][for
    details]{2010-AJ-140-Ryan}.
    \add{The mean albedo (last column) is obtained by averaging all the
      determinations, weighted by their accuracy and number.}
    \label{tab: albedo}
  }
\end{center}
\end{table*}

\subsection{Average density of each taxonomic class\label{sec: density}}

  \indent To convert from number of objects to
  mass, the average density for each class is crucial.
  Recently, an order of magnitude improvement of the sample of
  \add{asteroid} density estimates to 287 allowed the computation of the average density
  for each taxonomic class \citep{2012-PSS-73-Carry}. 
  In that work, multiple average densities are reported
  depending on the cutoff quality of measurements included.
  For the
  densities used in this work we chose the average densities using only the
  highest quality measurements (despite the smaller sample size).  While
  these values are certainly an improvement over assuming the same density
  for all asteroids, there is still significant uncertainty in the real
  densities for any single object, and there is likely a correlation between
  density and size particularly due to differences in macroporosity
  \citep[see][for details]{2012-PSS-73-Carry}.  
  Because we use such a large sample, the differences between
  any single asteroid and the average should have only a minor 
  effect on the outcome. \\
  \indent For E, M, and P class objects no average density was reported in
  \citet{2012-PSS-73-Carry}. In 
  this work we take all objects with densities in each of those classes and
  calculate average densities for each class. We find densities of
  $\rho_E$\,=\,2.8\,$\pm$\,1.2,
  $\rho_M$\,=\,3.5\,$\pm$\,1.0, and
  $\rho_P$\,=\,2.7\,$\pm$\,1.6\,g/cm$^3$. 
  The density of M-types is the highest which is consistent with the
  current interpretation for their composition.
  Some objects in that class are thought to be metallic,
  and to contain significant amounts of dense iron 
  \add{\citep[e.g.,][]{1989-AsteroidsII-Gaffey, 1989-AsteroidsII-Lipschutz,
      1989-AsteroidsII-Bell}. }  
  However, the M-class is
  degenerate in both visible spectrum and geometric albedo because multiple
  kinds of asteroids are known to fall in that category each having
  different composition and density
  \citep[see][among others]{2000-Icarus-145-Rivkin, 2008-Icarus-195-Shepard,
    2010-Icarus-210-Ockert-Bell}. 
  Not enough data are available to
  confidently distinguish the distributions of the different objects falling
  in the M-class so we group them together in this work. Additionally, as no
  density measurements are available for the D class, we assign an average
  density of 1\,g/cm$^3$, a density consistent with
    comets and transneptunian objects from the outer solar
    system \citep{2012-PSS-73-Carry}.


  \section{Observing Biases\label{sec: bias}}

    \indent Asteroid observations over visible wavelengths are subject to
    multiple biases, and the SDSS dataset is no exception. Detection
    biases for automatic surveys (relevant to discovery surveys as well
    as SDSS) \add{are due to} properties of the asteroid (such as size, albedo,
    and distance), the physical equipment (such as telescope size and CCD quality),
    the scan pattern of the sky,
    and the software's automatic detection algorithm.
    For a thorough description
    of asteroid observing biases see
    \citet{2002-AsteroidsIII-Jedicke}.\\
    \indent Efforts to correct 
    the observed asteroid distribution from observing
    biases have been \add{undertaken} for decades
    \citep{1971-NASSP-Chapman, 1971-NASSP-Kiang, 1982-Science-216-Gradie,
      1989-AsteroidsII-Gradie, 
      1999-PhD-Bus, 2004-Icarus-170-Stuart}.  One of the most 
    significant is a bias toward observing objects with the highest apparent brightness
    (objects that are larger, closer, or have a higher surface albedo). This
    bias is particularly important for the smallest asteroids, where the
    incompleteness of observed versus as-yet undiscovered asteroids is
    considerable for any magnitude-limited survey.\\
    \indent \add{The basis of relating the information in the given dataset to the entire 
      suite of asteroids done here is fundamentally the same as in most previous 
      work, however, it is executed slightly differently. In previous work 
      \citep[\eg,][]{1971-NASSP-Kiang, 1971-NASSP-Chapman, 1975-Icarus-25-Chapman,
        1982-Science-216-Gradie, 1985-Icarus-61-Zellner,
        1999-PhD-Bus, 2002-Icarus-158-BusI, 2003-Icarus-162-Mothe-Diniz} 
      the asteroid belt is broken 
      up into bins based on orbital elements (typically semi-major axis but some 
      works include inclination as well) and brightness (earlier works used
      the apparent magnitude in V but later works used the absolute magnitude H).
      A correction factor is calculated 
      as the total discovered numbered objects in each bin divided by the total number 
      of objects in each bin in the given dataset. Each object in the dataset is then multiplied 
      by the appropriate correction factor. }\\
    \indent \add{In this work we determine the fraction of each taxonomic class in each bin from 
      our dataset and apply those fractions to the total number of discovered objects. 
      These methods are most accurate if the original dataset is essentially an unbiased
      dataset and assume the relative fractions in each bin in the given dataset represent 
      the actual relative fractions of all asteroids. We describe in this work many steps to
      both minimize bias in the dataset and to most accurately compare objects of similar 
      size. These include using the average albedo per taxonomic class to move from an 
      H magnitude-limited to diameter -limited sample and correcting for discovery 
      incompleteness at large H magnitudes. This work accounts for both the sensitivity 
      difference between the inner and outer parts of the belt and uses a dataset sensitive 
      enough to probe to very small sizes.  } \\
    \indent Many of the previous spectroscopic surveys were subject to a target selection bias.
    These surveys focused more heavily on objects within asteroid families making the
    sample weighted more strongly toward these objects.  \add{Previous work
    included a correction for these biases \citep[\eg,][]{2002-Icarus-158-BusII, 
    2003-Icarus-162-Mothe-Diniz}}. However, because the SDSS is an \add{automated}
    survey that does not specifically target any type of objects or region of
    the belt it does not have the bias of many of the asteroid spectroscopic
    surveys that targeted specific regions. \\
    \indent It is also arguable that, even after correcting
    for this selection bias, counting family members overweights the
    importance of the original parent  
    body in terms of overall compositional distribution. 
    \add{Even with an ideal, unbiased dataset, if one counts each asteroid with an 
    equal weight (for example, by number) the compositional distribution 
    will be heavily weighted toward the asteroid families even though all the 
    family members are essentially of the same composition and 
    originate from the same body. This is fine for studies of number distributions, but
    not for the distribution of total material.
    A way to mitigate this oversampling of families is to explore the distribution 
    in terms of volume or mass as explained in the introduction. In this case we are
    counting all contributed material of the family; in essence we are putting the
    ejected fragments back together again and accounting for the total amount 
    of material. } \\
    \indent Accounting for the bias amongst the smallest asteroids is common to 
    many datasets. Unique to SDSS compared to previous spectroscopic work
    is the bias against observing the largest, brightest asteroids because they
    saturated the SDSS detector. Any study of the SDSS sample
    would need to correct for the missing large asteroids.

  \section{Defining the least-biased subset\label{sec: subset}}

  \subsection{Corrections for the largest, brightest asteroids}

    SDSS did not have the capability to measure the largest, brightest
    asteroids. Conveniently, past spectroscopic surveys are nearly complete at these sizes 
    and fill in that gap (Fig.~\ref{fig: sdsscomplete}).
    
    We include the taxonomic classes for 1,488 asteroids with an
    absolute magnitude H\,$<$\,12 determined using spectroscopic measurements in
    the visible wavelengths 
    \citep{1985-Icarus-61-Zellner, 
      2002-Icarus-158-BusI, 
      2004-Icarus-172-Lazzaro, 
      2009-Icarus-202-DeMeo}, available on the
    Planetary Data System \citep{PDSSBN-NEESE}. We
    keep only the large objects from these surveys where spectroscopic
    sampling is nearly complete ($>$90\%). The smaller objects in the spectroscopic 
    surveys (H\,$>$\,12) were not included in this work because
    they are more subject to observing biases and selection criteria
    \citep{2003-Icarus-162-Mothe-Diniz}.  
    \add{If an object was observed both in the spectroscopic surveys and the 
    SDSS dataset, we use the data and classification from the spectroscopic surveys.}

  \subsection{Corrections for the smallest, dimmest asteroids}

    Rather than extrapolating into regions in which
    we have no data that could misrepresent reality,
    we instead remove the biased portions of the data.
    \add{We determine
      the size of the smallest, dark asteroid
      at a far distance (in this case, the outer belt) at which the SDSS survey is 
      highly efficient. This number is based on the magnitude limits 
      given by \citet{2001-AJ-122-Ivezic} and the turnover in objects detected in the survey
      as a function of size (described in the next paragraph).} We then remove any
    asteroids from the sample that are smaller than that limit.      
    In essence, we create a sample restricted by a physical rather
    than an observable quantity: a \add{diameter}-limited instead of an apparent
    magnitude-limited sample.

      \indent Taking the SDSS sample, we determine the
      largest absolute magnitude (H) at which the survey is sensitive for each
      zone.
      We present in Fig.~\ref{fig: sdsscomplete} the number of objects
      and fraction of the sample covered by the spectroscopic surveys as well as
      the fraction the SDSS covers relative to all discovered and undiscovered
      asteroids for a large range of absolute magnitudes. 
      The peak of the black solid line in Fig.~\ref{fig: sdsscomplete}
      represents the limit of
      discovery efficiency for zones of the main belt. The cutoff magnitudes are
      roughly 17.2, 16.5, 15.5, 14.5, and 12.5 for the inner (IMB), middle
      (MMB), and outer Main Belt (OMB), Cybeles and Hildas, and Trojans,
      respectively. \\ 
      \indent We use these absolute magnitude limits to define the asteroid size range for
      which a distribution study can be reasonably confident. The smallest size
      sampled among all asteroid types is limited by the darkest, farthest
      objects (P-type, see Table~\ref{tab: cuts}).
      For our sample we use the outer main belt
      to determine our size cutoff. It would be preferable to use the Hilda or
      Trojan regions, because then we explore the same size range from the main
      belt out to the Trojans. However, this would drastically limit our
      sample size. It is thus important to recognize that our results do not
      contain Hildas and Trojans down to as small sizes as in the main belt. 
      In our sample, the number of Hildas and Trojans is severely biased toward 
      larger sizes, however, because these populations contain asteroids
      all with similarly low albedos
      \add{\citep{2011-ApJ-742-Grav, 2012-ApJ-744-Grav, 2012-ApJ-759-Grav}}
      there is no significant bias on the relative
      number of bodies of each taxonomic class. For this reason we include
      the Hildas and Trojans in the present work. \\
      \indent The smallest P-type asteroids the SDSS surveyed in the OMB have H=15.5
      which represents a diameter of $\sim$5\,km. While we sample, for
      example, S-types in the outer belt and C-types in the inner belt with
      diameters of $\sim$2\,km and S- and V-types in the inner belt to 1\,km or less,
      including these smaller objects in
      our sample would bias the results in terms of number toward
      these smaller objects that are not sampled in the outer belt. 
      Instead we
      include in our sample only objects that are 5 kilometers or larger. 
      This size is equivalent to a different H magnitude for each class.
      The ratio of each taxonomic class' albedo ($a_i$, where $i$ is the
      taxonomic class) with the P-type albedo ($a_P$) can be used to determine the
      magnitude difference between same-size objects of different taxonomic
      classes using the equation 
\begin{equation}
  H_i - H_P = 2.5 \log{ \frac{a_P}{a_i} }
\end{equation}

\begin{figure*}[t]
  \includegraphics[width=\textwidth]{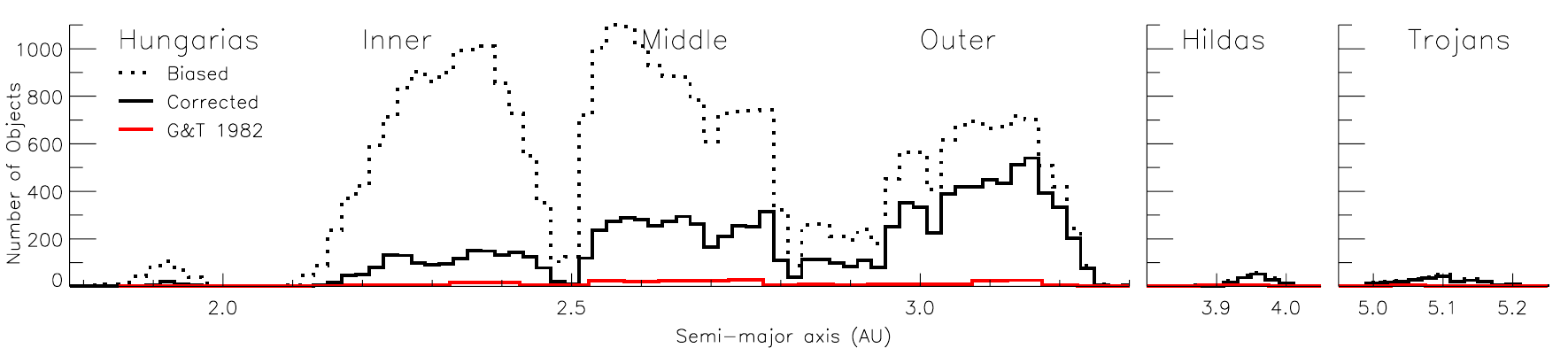}
  \caption[Comparison of 1982 and 2013 samples]{%
    Number of asteroids as a function of heliocentric distance for three different
    samples: our original sample made of spectroscopic surveys and SDSS
    photometry (34,503 asteroids, dashed line), our bias-corrected sample (13,211
    solid line), and the sample of 656 taxonomically classified asteroids from
    \citet{1982-Science-216-Gradie}.
    \label{fig: vs82}
  }
\end{figure*}

    We cut the sample of each taxonomic class according to these H magnitude
    limits, which are listed in Table~\ref{tab: cuts}.
    The average albedo for each class
    was determined by taking the average of the albedo determined for each
    class from IRAS, AKARI, and WISE
    \citep[][see Sec.~\ref{sec: albedo}]{
      2010-AJ-140-Ryan, 2011-PASJ-63-Usui, 2011-ApJ-741-Masiero}.
    Using a different H magnitude for each taxonomic class is
    critical. If we cut our sample at H=15.5 for all objects we would be
    comparing, for example, 5\,km P-types to 2\,km S-types\add{, which are
      much more numerous owing to the steep size-frequency distribution of
      the asteroid population}.\\
    \indent The size of the SDSS sample before and after the bias-correction
    selection is shown in Fig.~\ref{fig: vs82}, together with the number of
    objects presented in the preceding work by
    \citet{1982-Science-216-Gradie}.  It is clear that a vast
    number of objects are removed from the inner and middle sections of the
    belt because they are below the critical size limit.  To give an
    estimate on the importance of this size correction, there are 
    $\sim$5,000 5\,km asteroids in the middle belt, however there are
    about $\sim$40,000 2\,km ones known, nearly a factor of 10 greater.

\begin{table}
\begin{center}
\begin{tabular}{cc c@{\,$\pm$\,}cc@{\,$\pm$\,}c}
\hline
\hline
Class & H$_{\textrm{\small cut}}$ & 
  \multicolumn{2}{c}{Density} &
  \multicolumn{2}{c}{Albedo } \\
\hline 
A & 13.99 & 3.73 & 1.40 & 0.20 & 0.03 \\
B & 14.38 & 2.38 & 0.45 & 0.14 & 0.04 \\
C & 15.30 & 1.33 & 0.58 & 0.06 & 0.01 \\
D & 15.30 & 1.00 & 1.00  & 0.06 & 0.01 \\
K & 14.38 & 3.54 & 0.21 & 0.14 & 0.02 \\
L & 14.46 & 3.22 & 0.97 & 0.13 & 0.01 \\
S & 13.84 & 2.72 & 0.54 & 0.23 & 0.02 \\
V & 13.39 & 1.93 & 1.07 & 0.35 & 0.01 \\
E & 13.12 & 2.67 & 1.20 & 0.45 & 0.21 \\
M & 14.49 & 3.49 & 1.00 & 0.13 & 0.05 \\
P & 15.50 & 2.84 & 1.60 & 0.05 & 0.01 \\
\hline 
\end{tabular}
  \caption[Absolute magnitude, density, albedo per taxonomic class]{%
    Cuts on the absolute magnitude for each taxonomic class. \add{These cutoffs
    were determined by the limiting case of P-type asteroids in the outer belt.}
    Average density \citep[in\,g/cm$^3$, from][]{2012-PSS-73-Carry}
     and albedo
    (see \ref{sec: albedo})
    are also reported. 
    We choose a density of D-types of 1\,g/cm$^3$,
    consistent with an outer solar system origin because
    no D-type densities have been accurately measured.
    \label{tab: cuts}
  }
\end{center}
\end{table}

\section{Taxonomic classification\label{sec: taxo}}

  \indent The SDSS asteroid data has been grouped and classified according
  to their colors by many authors. \citet{2002-AJ-124-Ivezic} classified the C, S,
  and V groups using the z'-i' color and the first principal component of the
  r'-i' \textsl{vs} r'-g' colors.
  \citet{2005-Icarus-173-Nesvorny} used the first two principal components of
  u', g', r', i', z' colors 
  and distinguished between the C, X, and S-complexes. \citet{2010-AA-510-Carvano} converted
  colors to reflectance values and created a probability density map of previously classified
  asteroids and synthetic spectra to classify the SDSS dataset.
     
  \indent In this work we seek to
  maximize the taxonomic detail contained in the dataset and strive to keep
  the class definitions as consistent as possible with previous spectral
  taxonomies that were based on higher spectral resolution and larger
  wavelength coverage data sets, 
  specifically Bus \citep{2002-Icarus-158-BusII}
  and Bus-DeMeo \citep{2009-Icarus-202-DeMeo}
  taxonomies.

  \subsection{Motivation for manually defined class boundaries}
      \indent The best way to mine the most information out of such a large
      dataset could be to perform an analysis of the variation and
      clustering. Methods such as Principal Component Analysis or Hierarchical
      Clustering could separate and highlight groups within the data. The advantage to
      automated methods is they are unbiased by human intervention and can
      efficiently characterize large datasets, which are the motivations for
      many unsupervised classifications.  \\
      \indent However, because most of our understanding of asteroid mineralogy comes
      from relating asteroid spectral taxonomic classes to meteorite classes
      and comparing absorption bands, we find it more relevant to \add{connect} this
      low-resolution data to already defined and well-studied asteroid
      taxonomic classes (that were based on Principal Component
      Analysis). This facilitates putting the SDSS results in context
      with the findings from other observations that have accumulated over
      decades. To classify the data we started with the class centers and
      standard deviations \add{(based on data used to create the Bus-DeMeo taxonomy
      converted to SDSS colors)}  to
      calculate the distance of each object to the class center. \\
      \indent Considering the above, while we still use the class centers and
      calculated deviations as a guide, we choose to fix boundaries for each
      class and manually tweak them (as described below) according to the data
      to best capture the essence of each class. A negative consequence of
      fixed boundaries is that near the boundary objects exist on either side
      that may have very similar characteristics though are classified
      differently (as opposed to methods which assign a probability for each
      object to be in a certain class). Additionally, a human bias is
      added. The advantage, however, is we are forced to carefully evaluate
      the motivation for the definition of each class to group objects
      according to the most diagnostic spectral parameters (particularly
      considering the much wider spread of the SDSS dataset), consistency with
      previous classifications, and potential compositional
      interpretation. Additionally, fixing the boundary allows us to more easily use the
      classifications as a tool. We can use these
      classifications to determine the fraction of objects in each class and
      the mass of each taxonomic type across the solar system.

  \subsection{Defining the class boundaries}
    \indent We transform the apparent magnitudes from SDSS to reflectance values to
    directly compare with taxonomic systems based on reflectance data. We then subtract
    solar colors in each filter and calculate reflectance values using the
    following equation: 
\begin{equation}
R_f = 10^{-0.4 \left[ 
                \left( M_f - M_g \right)
               -\left( M_{f,\odot} - M_{g,\odot} \right)
              \right]}
\end{equation}
    
    \noindent where ($M_f$) and ($M_{f,\odot}$) are the magnitudes of the
    object and sun in a certain filter $f$, respectively, at the central
    wavelength of the filter. The equation is normalized to unity at the
    central wavelength of filter g using
    ($M_g$) and ($M_{g,\odot}$): the $g$ magnitudes of
    the object and sun, respectively.
    Solar colors used in this work are
    r'-g'= -0.45\,$\pm$\,0.02,
    i'-g'= -0.55\,$\pm$\,0.03, and
    z'-g'= -0.61\,$\pm$\,0.04 from \citet{2006-MNRAS-367-Holmberg}.
    Note that because we use solar colors in the Sloan filters we do
    not convert from the g', r', i,' z' filters (central wavelengths:
    g'=0.4686, r'=0.6166, i'=0.7480, z'=0.8932\,$\mu$m)
    to standard g, r, i, z filters. As mentioned in Section~\ref{sec: selectSDSS},
    we do not use the u' filter because of the very large errors for this datapoint.
    
    \indent The classification of the dataset is based on two dimensions:
    spectral slope over the g', r', and i' reflectance values (hereafter gri-slope),
    representing the slope of the continuum, and z'-i' color, representing
    band depth of a potential 1\,$\mu$m band. We restrict the evaluation of the
    spectral slope to g', r', and i' filters only, excluding the z' filter
    because it may be affected by the potential 1\,$\mu$m band. These two
    parameters (slope and band depth) are the most characteristic spectral
    distinguishers in all major taxonomies beginning with
    \citet{1975-Icarus-25-Chapman}
    because they account for the largest amount of meaningful and readily
    interpretable variance in the system. 

    \indent We choose not to use \astar~defined by
    \citet{2002-AJ-124-Ivezic} or the first Principal
    Component (PC1) defined by
    \citet{2005-Icarus-173-Nesvorny} used in other works. 
    \add{$a^\star$ is the first principal component of the
    r'-i' versus g'-r' colors and PC1 is
    the first principal component of the measured fluxes
    of all five filters. To most effectively use Principal
    Component Analysis, the dimension with the greatest
    variance, slope in this case, should be removed before
    running PCA
    to increase sensitivity to more subtle variation
    \citep[see discussion in][]{1999-PhD-Bus}. We also disfavor 
    the inclusion of the u' filter (used for PC1) as it adds significant noise 
    to the data (Fig~\ref{fig: filters}).
    We find our slope parameter is reasonably 
    well-correlated with \astar~but not well-correlated with PC1, as expected
    from the use of u' photometry in PC1.}

\begin{figure}[t]
  \includegraphics[width=.5\textwidth]{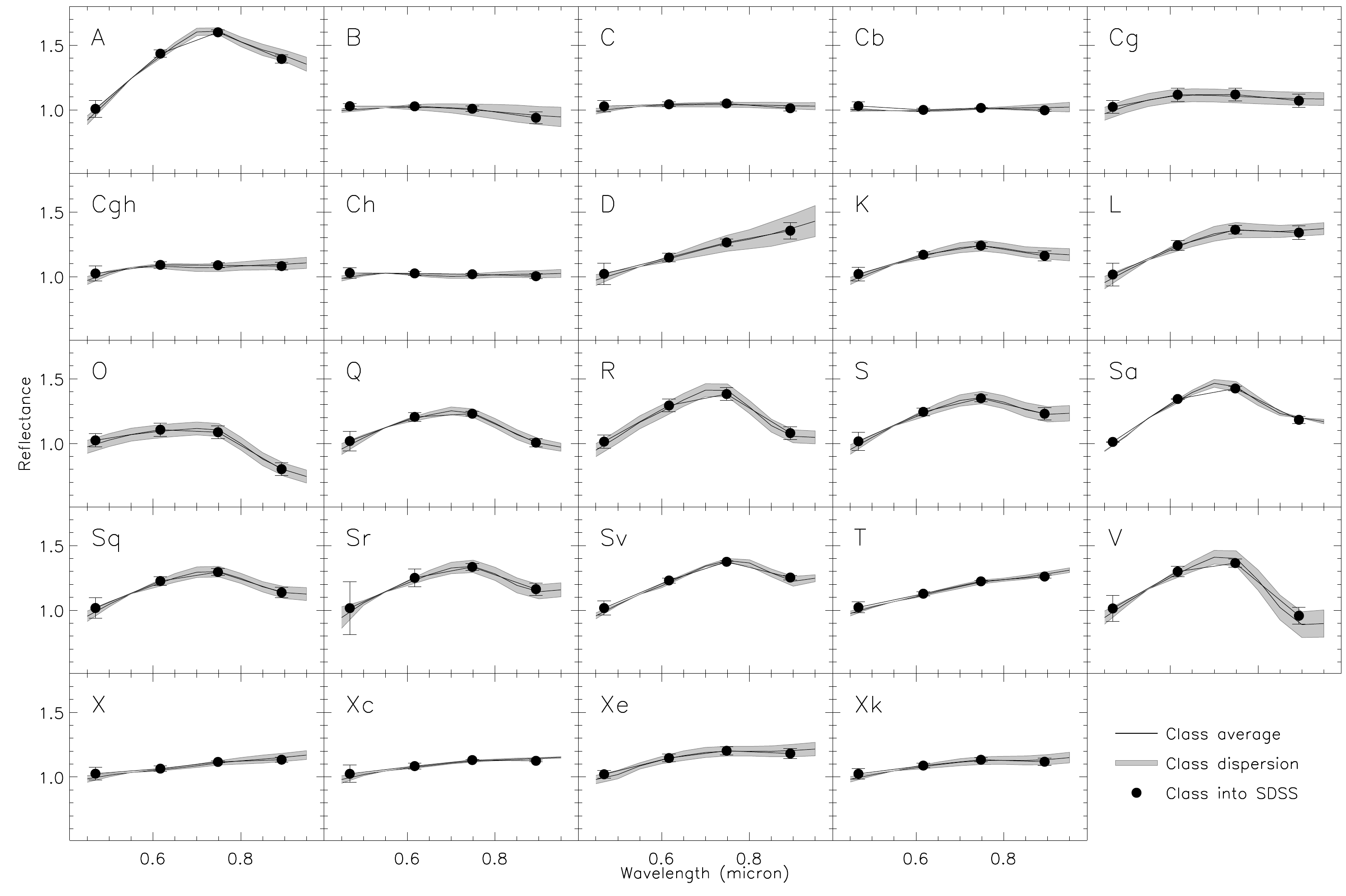}
  \caption[Average SDSS colors of Bus-DeMeo spectra]{%
    Average Bus-DeMeo \citep{2009-Icarus-202-DeMeo}
    spectra converted to SDSS colors used to define the
    classification boundaries. The black dots (with 1 standard deviation from
    the mean plotted) represent the average Bus-DeMeo spectra converted into
    SDSS colors. The u' filter is extrapolated from the data because the
    spectra do not cover those wavelengths (the u' filter is not used in the
    classification of SDSS data, however). The gray background plots the
    average spectrum plus one sigma for comparison with the colors. Because
    the Cg, O, and R classes are defined by a single object the standard
    deviation is set to 0.1. 
    \label{fig: bdt}
  }
\end{figure}

    \indent \add{We base the classification
    on the 371 spectra used to create} the Bus-DeMeo taxonomy
    \citep{2009-Icarus-202-DeMeo},
    whose classes are very similar those of the Bus taxonomy
    \citep[][]{2002-Icarus-158-BusII}, with a few
    classes removed.
    The variation among the reflectance spectra of the 371
    asteroids used to define the Bus-DeMeo classes helped guide the boundary
    conditions of the present SDSS taxonomy. We convert all the spectra into
    SDSS reflectance values by convolving them with the SDSS filter
    transmission curves\footnote{
      \href{http://www.sdss.org/dr7/instruments/imager/}{http://www.sdss.org/dr7/instruments/imager/}},
    thus providing the average SDSS colors and standard
    deviation per class (see Fig.~\ref{fig: bdt})

%
      \indent Because the SDSS data have a spectral resolution
      significantly lower than the Bus-DeMeo data set
      (see Fig.~\ref{fig: bdt}) and subtle
      spectral details are lost, we combine certain classes into their broader
      complex. The C-complex encompasses the region including C-, Cb-, Cg-,
      Cgh-, and Ch-types. The S-complex encompasses the S-, Sa-, Sq-, Sr-, and
      Sv-types. The X-complex includes X-, Xc-, Xe-, Xk-, and T-types. The
      classes that are maintained individually are A, B, D, L, K, Q, and V.
      While we distinguish all these classes based on the SDSS colors here, we
      slightly modify our use of some of these classes for this work
      (see Section~\ref{sec: mods}).
      We do not classify the rare R- or O-type in this
      work, because there is significant overlap between O-types or R-types
      and other classes in the visible wavelength range, and they are
      particularly rare classes. The R class would
      overlap the V class essentially spanning the shallower z'-i' ``band depth''
      region. We tested separating the R class, but the majority of the
      objects classified as R were located in the Vesta family.  

\begin{figure}[t]
  \includegraphics[width=.5\textwidth]{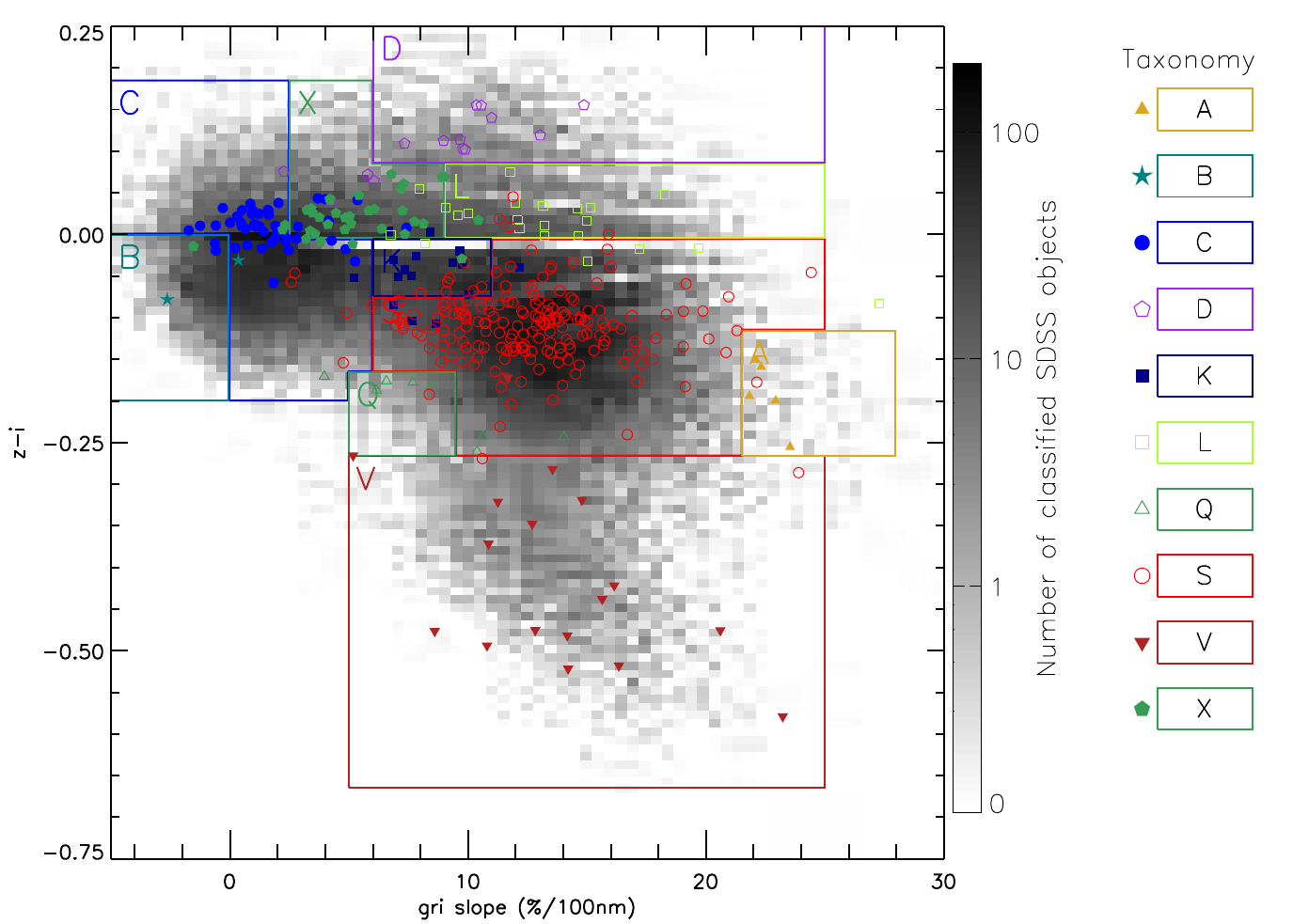}
  \caption[Classification boundaries]{%
    Boundaries used to classify SDSS data into taxonomic classes. The colored
    points are the spectra from the Bus-DeMeo taxonomy
    \citep{2009-Icarus-202-DeMeo} converted to SDSS
    colors. In the background, the density of the number of objects from MOC4 are
    plotted to show the dispersion of the SDSS data. 
    \label{fig: boundary}
  }
\end{figure}

      While the Bus-DeMeo class averages are very useful as a guide, the
      system was based on a sample size three orders of magnitude smaller than
      present SDSS sample. The SDSS dataset therefore shows a much more
      continuous range of reflectance characteristics. To compare the two
      datasets, we plot the distribution of SDSS objects in z'-i' color and
      gri-slope, with the 371 objects from the Bus-DeMeo taxonomy
      (Fig.~\ref{fig: boundary}).
      Furthermore, the figure shows the boundaries for each class
      defined in this work. We drew boundaries that best separated each class
      based on the position of the class centers and standard deviations
      \add{based on the 371 spectra dataset.} We
      visually inspected each boundary by plotting the spectral data on each
      side of the boundary and comparing them with the designated class to
      tweak the position of the line and best separate each class. 

      We strove to preserve the uniqueness of the more exotic classes,
      restricting A- and D-types to the outliers with the largest slopes, and
      Q- and V-types with the deepest bands. The B-type was defined to have
      both a large, negative gri-slope and a negative z'-i' value. A list of all
      the boundaries is provided in Table~\ref{tab: bound}.
      Classification is performed in
      decision tree form, where the gri-slope and z'-i' value of the asteroid is
      compared with each region in the following order: C, B, S, L, X, D, K,
      Q, V, A. If the object falls in more than one class, it is designated to
      the last class in which it resides. As can be seen in
      Fig.~\ref{fig: boundary}, there
      are a handful of objects that reside outside the defined classes. We
      give these objects the designation ``U'', historically used to mark
      unusual objects in a sample that do not fall near any class. We do not
      include these objects in our study. Most of these extreme behaviors are
      likely due to problems with the data even though no flags were
      assigned (see details in Section~\ref{sec: selectSDSS}).
      Follow-up observations could determine whether the objects
      really are unique.

\begin{table}[ht]
\begin{center}
\begin{tabular}{crrrr}
\hline
\hline
Class & \multicolumn{2}{c}{Slope  (\%/100 nm)}& \multicolumn{2}{c}{z'-i'} \\
 & (min) & (max) & (min) & (max) \\
\hline
A &  21.5 &  28.0 & -0.265 & -0.115 \\
B & -5.0 & 0.0 & -0.200 &  0.000 \\
C & -5.0 &  6.0 & -0.200 &  0.185 \\
D &  6.0 &  25.0 &  0.085 &  0.335 \\
K &  6.0 &  11.0 & -0.075 & -0.005 \\
L &  9.0 &  25.0 & -0.005 &  0.085 \\
Q &  5.0 &  9.5 & -0.265 & -0.165 \\
S &  6.0 &  25.0 & -0.265 & -0.005 \\
V &  5.0 &  25.0 & -0.665 & -0.265 \\
X &  2.5 &  9.0 & -0.005 &  0.185 \\
\hline
\end{tabular}
  \caption[Boundaries of the taxonomic classification]{%
    Table of classification boundaries. The classification is performed
    following the order: C, B, S, L, X, D, K, Q, V, A.
    \label{tab: bound}
  }
\end{center}
\end{table}

  \subsection{Determining a single classification for multiple observations}
    \indent Of the many observations in the SDSS MOC that remained after we
    applied cuts on the photometric precision
    (58,607, see Section~\ref{sec: selectSDSS}), many were actually the
    same object observed more than once. The number of unique objects in our sample is
    34,503. \add{For some of these objects, not all observations fell into the same class.}
    Because we seek to categorize each object with a unique class,
    we use the following criteria to choose a single
    class for any object that has multiple observations that fall under
    multiple classifications (5,401 asteroids, \ie, 15.7\% of the sample):

    \begin{itemize}

      \item The class with the majority \add{number of classifications} is
        assigned (2,619 asteroids, \ie, 
        7.6\% of the sample) 
      \item If two classes have equal frequency and one of them is C, S,
      or X we assign the object to C, S, or X, continuing the philosophy
      of remaining conservative when assigning a more rare class (1867
      asteroids, \ie, 5.4\% of the sample) 
      \item If the two majority classes are C/S, X/C, S/X (or three competing
      classes of C/S/X) we assign it to the U class and disregard those objects
       in the distribution work (919
      asteroids, \ie, 2.6\% of the sample). We prefer to keep the sample
      smaller, rather than contaminate it with objects that we have randomly
      chosen a classification among C, S, or X and thus possibly bias the
      sample.  
      
      \item \add{For objects that are assigned multiple classes but none is either 
      the majority or C, S, or X is assigned to the U class.}
    \end{itemize}

    \indent Among the largest asteroids, particularly those between H magnitudes 
    of 9 and 12, several asteroids observed by
    the SDSS had taxonomic classes from previous spectroscopic measurements. 
    The classification based on SDSS and previous work were generally consistent,
    but in cases that differed, we assigned the asteroid to the class determined by
    spectroscopic measurements.

  \subsection{Caution on taxonomic interpretation \label{sec: caution}}
    \indent One must be careful when interpreting the classifications
    presented here. First, the resolution of the SDSS data are significantly
    lower than the spectra to 
which they are compared. Second, the fact that
    we find multiple classifications for multiply observed objects suggests
    there is a larger uncertainty in the data than expected. Third, for many
    classes (particularly L, S, Q, A), the visible data can only suggest the
    presence of a 1\,$\mu$m band, but do not actually predict the depth or shape of
    that band
    \citep[for more detail see][]{2009-Icarus-202-DeMeo, 2010-PhD-DeMeo}.
    This is important because,
    for example, a spectrum might look closer to a K- or an L-type in the
    visible range, but near-infrared data could place them more confidently in
    the S-class (or vice versa).  

    Each class is meant to be representative of a certain spectral
    characteristic, but with limited wavelength coverage and limited
    resolution, there is some degeneracy. For example, the Q-class defined
    here represents objects with a low slope and moderate 1\,$\mu$m band depth. We
    do not suppose that all objects classified as a Q-type are young, fresh
    surfaces as is typically associated with the Q class. Careful follow-up
    observations are important to make such a claim.

    Defining boundaries for C, X, and D-types is not easy because they are
    distinguished only by slope and there is a continuous gradient of slope
    characteristics. This problem is not unique to the SDSS dataset. The
    boundary between each type is somewhat arbitrary. The difference between a
    C-type of slope zero and a D-type with a high slope is meaningful, however
    we do not yet know how to interpret the significance of these spectral
    differences. It is likely that there is some contamination between C- and
    X-types with our classification scheme, though it is unlikely that much
    contamination exists for example between C- and D-types that are more
    easily distinguished.

  \subsection{Verification of our classifications \label{sec: verif}}

\begin{figure}[t]
  \includegraphics[width=.5\textwidth]{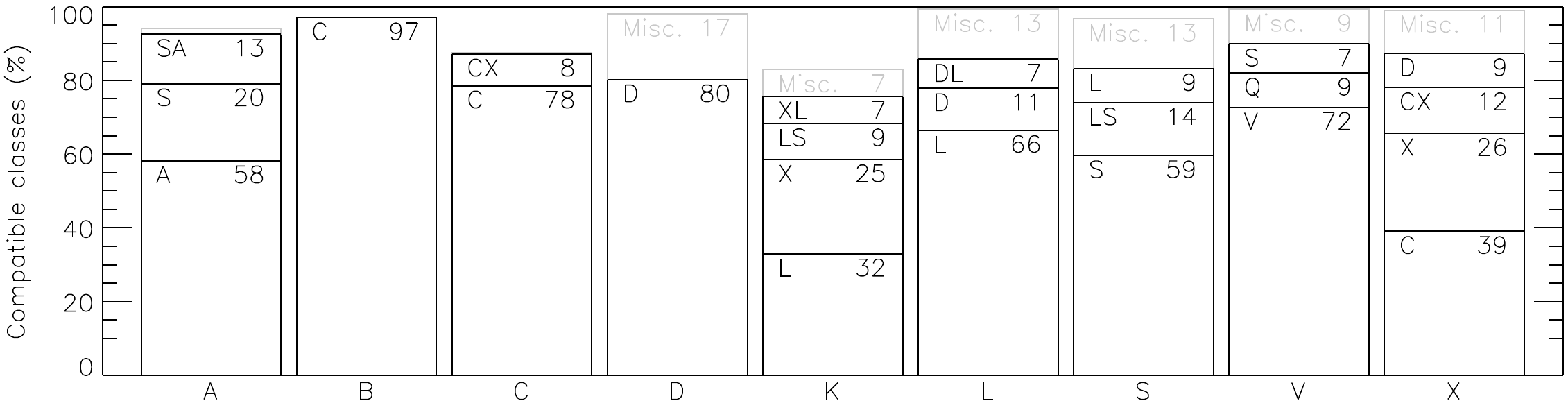}
  \caption[Comparison with Carvano classifications]{%
    Comparison of classifications in this work to those of \citet{2010-AA-510-Carvano}.
    For each class in our work a bar represents how those objects are 
    classified in the Carvano system. Some objects were given two letters by
    Carvano given in the PDS archive \citep{PDSSBN-Hasselmann}.
    We categorize according to the most numerous classes assigned by Carvano
    compared to this work. All ``compatible'' classes are shown since they are 
    relatively in agreement based on classes that border others. These highlight
    the small but compatible differences between the classifications. 
    Miscellaneous includes other classes we feel are compatible but make up
    a small percentage of the sample. All B-types 
    in our work are classified as C-type by Carvano because they do not make a 
    distinction between the two. The small unlabeled fraction
    represents mismatches
    where our work and Carvano's get significantly different results. The right side
    of each bar labels the percent of the total each Carvano class represents.
    \label{fig: carvano}
  }
\end{figure}

    \indent \add{With a unique class assigned to each object in our dataset
    we can now evaluate the robustness of our classifications.
    First, we compare the classification of each asteroid with the
    results of \citet{2010-AA-510-Carvano} available on the Planetary
    Data System \citep{PDSSBN-Hasselmann}.
    Because their classification is based on the same dataset, it is not an
    entirely independent check.
    However, their classification method is 
    different so consistency between the two supports both
    methods. Fig.~\ref{fig: carvano} graphically compares
    the two classifications. We list the classification differences
    that are generally compatible but represent the different 
    choices each method made. We find the two classifications 
    quite consistent. Of the major classification differences between
    the two methods we suspect some are due to boundary condition differences
    and others are 
    due to Carvano's inclusion of the u' filter, which we exclude
    in our work (see Sec.~\ref{sec: selectSDSS}).}
    
\begin{figure}[t]
  \includegraphics[width=.5\textwidth]{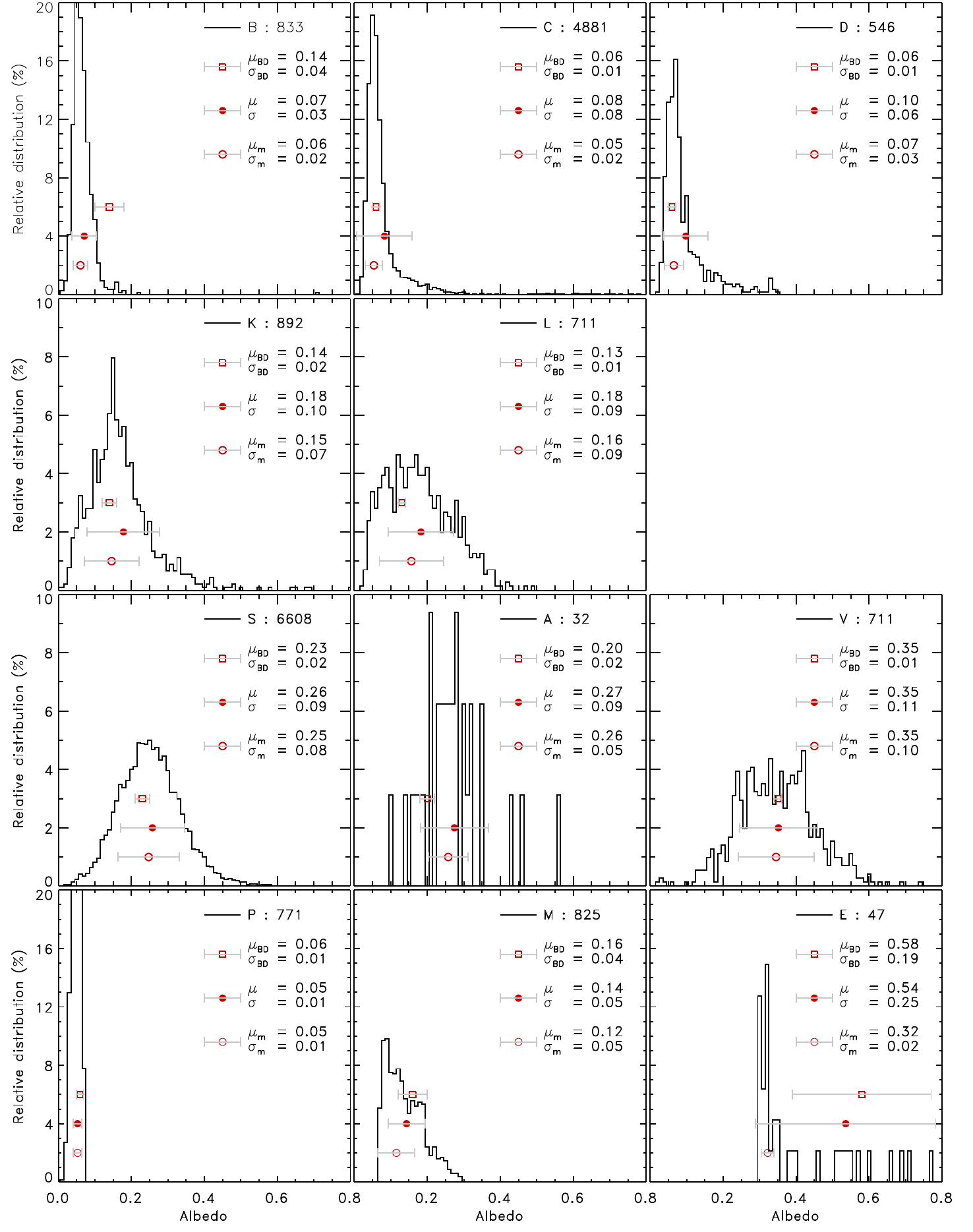}
  \caption[Distribution of albedo per class]{%
    Relative distribution of albedo for each class.
    For each class, we report the number of asteroids with albedo estimates,
    and the average albedo with its standard deviation ($\mu$, $\sigma$, open circle),
    together with the mode of the histogram
    ($\mu_m$, $\sigma_m$, filled circles).
    We also report the average albedo of the asteroids in the Bus-DeMeo sample 
    ($\mu_{\textrm{\tiny BD}}$, $\sigma_{\textrm{\tiny BD}}$, square symbol, see
    Table~\ref{tab: albedo}).
    \label{fig: albedo}
  }
\end{figure}

    \indent \add{Second, we retrieved the albedo of the asteroids as determined
    from IRAS, AKARI, and WISE data
    \citep{2002-AJ-123-Tedesco-a, 2010-AJ-140-Ryan, 
      2011-PASJ-63-Usui,
      2011-ApJ-741-Masiero, 2012-ApJ-759-Masiero, 
      2011-ApJ-742-Grav, 2012-ApJ-744-Grav, 2012-ApJ-759-Grav}. 
    We found 17,575 asteroids (out of 34,503, \ie, 51\%) with albedo
    determinations. We present in Fig.~\ref{fig: albedo} the distribution of
    albedo for each class, and the average values in Table~\ref{tab: albcheck}.
    The agreement between the average albedo per Bus-DeMeo class from previous
    work (see Table~\ref{tab: albedo}) and of the
    asteroids classified from SDSS colors gives confidence in our capability
    to assign a relevant class to these asteroids.
    One of the greatest differences is the albedo
    of the B class. }\\
%
\begin{table}[t]
\begin{center}
\begin{tabular}{crc@{\,$\pm$\,}cc@{\,$\pm$\,}c}
\hline
\hline
Class & N$_{objects}$&
    \multicolumn{2}{c}{Average} & 
    \multicolumn{2}{c}{Mode   } \\
\hline

A &	32    &  0.274  &  0.093  &  0.258  &  0.055	\\
B &	833   &  0.071  &  0.033  &  0.061  &  0.021	\\
C &	4881  &  0.083  &  0.076  &  0.054  &  0.023	\\
D &	546   &  0.098  &  0.061  &  0.065  &  0.026	\\
K &	892   &  0.178  &  0.099  &  0.146  &  0.075	\\
L &	711   &  0.183  &  0.089  &  0.157  &  0.088	\\
S &	6565  &  0.258  &  0.087  &  0.247  &  0.084	\\
V &	711   &  0.352  &  0.107  &  0.345  &  0.104	\\
E &	47    &  0.536  &  0.247  &  0.322  &  0.016	\\
M &	825   &  0.143  &  0.051  &  0.115  &  0.051	\\
P &	771   &  0.053  &  0.012  &  0.053  &  0.012	\\
\hline
\end{tabular}
  \caption[Average albedo per taxonomic class]{%
    Average albedo of each class based on the 17,575 objects
    in our SDSS dataset that had calculated albedos (51\% of our dataset).
    The results are consistent with previous albedo averages
    (Tables~\ref{tab: albedo} and~\ref{tab: cuts}) 
    strengthening the robustness of this work.
    \label{tab: albcheck}
  }
\end{center}
\end{table}
%
   \indent \add{We have separated the spectra of
    objects with negative slopes into the B class using SDSS colors as has 
    traditionally been done in taxonomic systems. This
    separation can be a useful indicator of spectral differences between B and
    C classes. The average albedo for B-types classified from spectroscopic
    samples is significantly higher than for C-types (see Table~\ref{tab: albedo})
    suggesting a compositional difference. 
    However, the average albedo of B-types in our SDSS sample is similar 
    to that for C-types so we caution that the B-types classified from
    spectroscopic surveys may not be fully representative of the B-types
    in our sample. }

\section{Building the compositional distribution\label{sec: distribuild}}

  \subsection{Additional taxonomic modifications\label{sec: mods}}
    \indent Keeping in mind the cautions mentioned in Section~\ref{sec: caution},
    for the taxonomic distribution work presented here
    we apply slight modifications to the
    classes. First, we note a significant over abundance of S-types in the Eos
    family. This is due to the similarity of S- and K-type spectra
    using only a few color points and the visible-only wavelength range. We
    thus reclassify all S-type objects to K-type within the Eos family 
    (defined by the family's current orbital elements
    a $\in$ [2.95, 3.1], 
    i $\in$ [8\degr, 12\degr], and 
    e $\in$ [0.01, 0.13]).
    Reviewing this change shows that the background of S-types
    is now evenly distributed, no longer showing a concentration within the
    Eos family. Additionally, for this study we group Q-types with the S-types
    because they are compositionally similar \citep{2010-Nature-463-Binzel,2011-Science-333-Nakamura}. \\
    \indent \add{In the previous section we discussed the albedo differences
    between B-types in our sample and B-types from other work.
    While future work may want to focus specifically on objects
    with negative slopes,  in this we choose to merge B-types
    with C-types classified by the SDSS dataset.} \\
    \indent \add{Our SDSS observations classify some Hildas and Trojans as K- and L-types.
      Careful examination reveals that for the K- and L-
      type objects that are near the border the X and D classes, the spectra could also be
      consistent with X and D. These Hilda and Trojan K- and L-types that have multiple observations are
      also classified X and D.} For example, the Centaur (8405)
    Asbolus has a very red visible spectral slope
    \citep{1999-AJ-117-Barucci}, categorizing it as a D-type. 
    Eight SDSS observations place this object in the L class and four
    in the D class. This difficulty is partially due to the degeneracy of the
    visible wavelength data. The Bus Ld class
    that is intermediate between the
    L and D classes does not remain an intact definable class when
    near-infrared data are available
    \citep{2009-Icarus-202-DeMeo}. There are four SDSS L-type Hildas and
    Trojans with albedos all of which are below 0.08 further suggesting that
    these objects are not characterized by what the K and L classes are
    compositionally meant to represent.
    \add{We therefore choose the 
      more conservative option to reclassify
      the Hilda and Trojan K- and L-types. The K-types that have slopes 
      more consistent with the X class are relabeled as X, while the L-types
      have slopes more consistent with D-type and are relabeled as D.} \\
    \indent Among the Hungarias, a population of small (H\,$>$\,13) C-types is
    seen. Upon 
    closer inspection, all (8) of the small C-types with WISE data have
    extremely high albedos (0.4--0.9), suggesting they are actually
    E-types (the high albedo group of X-types). This is unsurprising, as the
    Hungaria region is known to contain 
    a large population of high albedo E-type asteroids. We thus correct our
    Hungaria sample by assuming all small C-types are incorrectly classified,
    and remove those with H magnitudes greater than the E-type cutoff from our
    sample. \add{We expect some overlap between C-types and X-types (E, M, P)
    in other regions of the belt as well (as addressed in Sec.~\ref{sec: caution}) 
    although the classification of X v. C should be more balanced.}\\
    \indent \add{While we make these modifications for the objects in the SDSS sample 
    we do not make changes to the large objects classified spectroscopically from
    previous work. }

  \subsection{Discovery completeness \label{sec: complete}}
    
   \indent \add{While we select the subset of SDSS data where
   the survey is efficient, the dataset not complete. The information
   from the SDSS dataset must be applied to all existing asteroids in the 
   same size range. The Minor Planet Center (MPC) catalogues 
   all asteroid discoveries. Here we asses discovery completeness.}   

\begin{figure}[t]
\begin{center}
  \includegraphics[width=.5\textwidth]{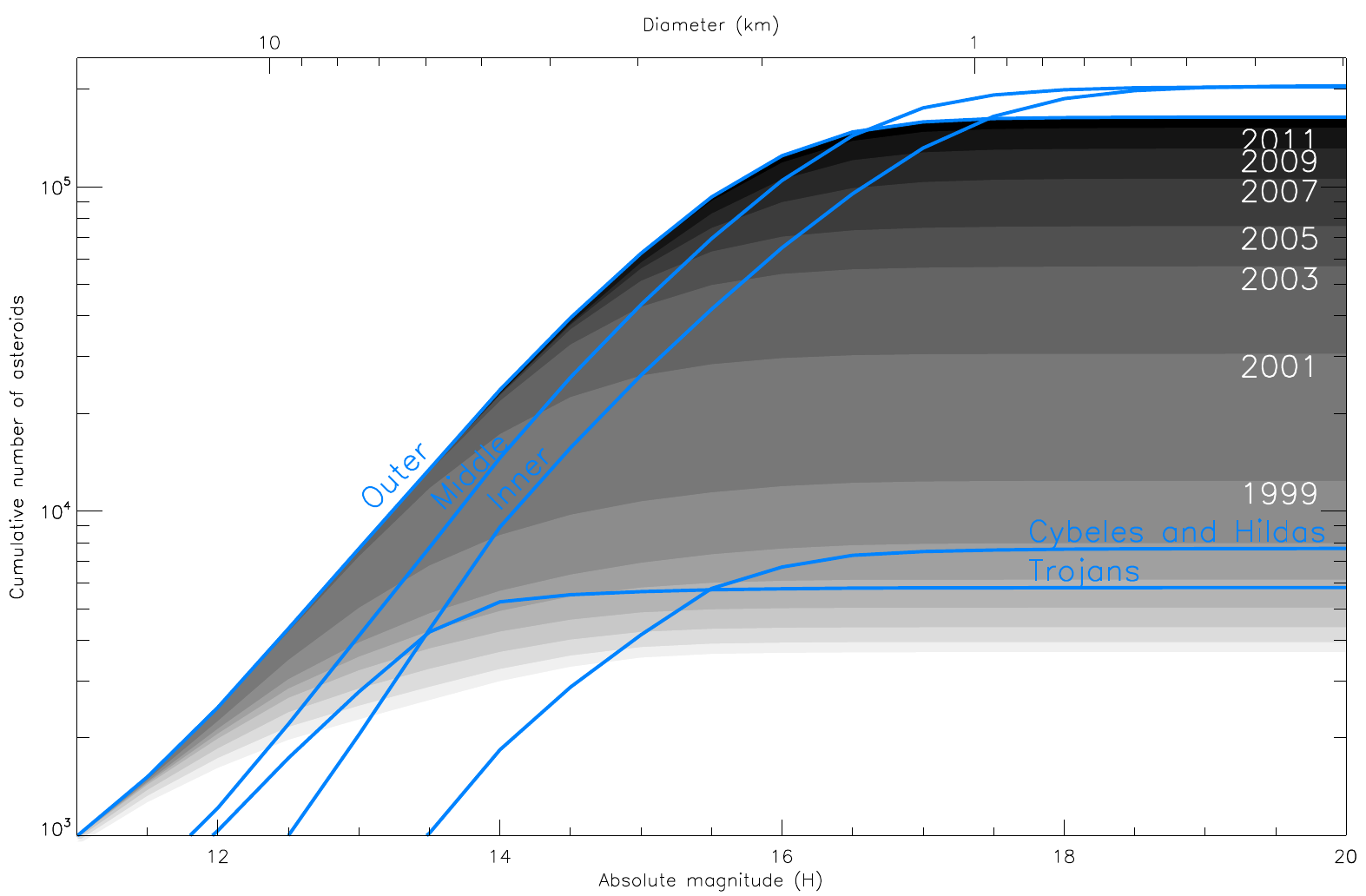}
  \caption[Completeness of minor planet discoveries]{%
    Discovery completeness through 2013. \add{For the outer belt, we plot the
    cumulative distribution as function of time up to 2013 January 01 (shades of gray),
    showing the evolution of the completeness limit to smaller
    (higher H) objects.
    We derive a limiting magnitude for the completeness of the
    MPC database of H\,=\,16, 15 and 14.5 (diameter of about 2, 3, and 4 km) for
    the inner, middle, and outer belt, respectively. }
    \label{fig: discomplete}
  }
\end{center}
\end{figure}

   \indent In Fig.~\ref{fig: discomplete}, we plot the cumulative number of
    discoveries in the outer belt for every two years of the past 10 years (to
    2013-01-01). 
    We derive a limiting magnitude for the completeness of the
    MPC database of H\,=\,16, 15 and 14.5 (diameter of about 2, 3, and 4\,km) for
    the inner, middle, and outer belt, respectively. 
    \add{We determine the completeness of small asteroids in each section of
      the main belt by extrapolating the size of the population using a power
      law fit to each region of the main belt
      (shown in Fig.~\ref{fig: sdsscomplete}).
      The difference between the currently observed populations and the extrapolated
      populations derived from these power laws provide the expected number of
      asteroids to be discovered at each size range}.
    The power law indices we find for the
    IMB, MMB, and OMB (determined over the H magnitude range
    14--16,
    13--15, and
    12--14.5) are -2.15, -2.57, and -2.42, respectively. These power law indices
    agree with other fits to the observations
    \citep{2009-Icarus-202-Gladman} as well as with the
    theoretical index calculated assuming a collision-dominated
    environment \citep{1969-JGR-74-Dohnanyi}.\\
    \indent For almost all H magnitudes
    in our sample we are nearly discovery complete. For the smallest size
    we use a power law function to determine completeness.
    In the H=15-16 magnitude range, we are 
    100\%, 85\%, and 60\% complete in the IMB, MMB, and OMB respectively.
    When applying the
    taxonomic fractions to the MPC sample of known asteroids we add 
    \add{a correction factor to account the 15\% and 40\% of objects that 
    have not been discovered in the middle and outer belt in the H=15-16 range.}
    For Cybeles, Hildas, and Trojans we do
    not extrapolate to determine sample completeness because there is far too much
    uncertainty in the size distribution of those populations due to fewer
    discoveries. We have not corrected these populations.
    The completeness of our dataset can be evaluated on 
    Fig.~\ref{fig: sdsscomplete}.\\
    \indent \add{There are undoubtedly still many objects yet to be discovered, especially
      at sizes smaller than we cover in this work. For reference, we explore the 
      total mass these undiscovered objects are expected to represent. The 
      largest objects represent the overwhelming majority of the mass in the 
      main belt. In fact, the asteroids from the spectral surveys (particularly H$<$10)
      represent 97\% of the mass 
      \citep[assuming a mass of 30\,$\times$\,10$^{20}$ kg for the entire main belt,
      ][]{2013-Icarus-222-Kuchynka}.}  \\
    \indent \add{We calculate the undiscovered mass \citep[assuming a general
        density of 2.0 \gcm~and an albedo of 0.18, 0.14, and 0.09 for the
        inner, mid, and outer main belt, based on WISE
        measurements, see][]{2011-ApJ-731-Mainzer} up  
      to H magnitude of 22 to be
      5.7\,$\times$\,10$^{12}$,
      4.8\,$\times$\,10$^{13}$, and
      1.6\,$\times$\,10$^{14}$\,kg for the 
      IMB, MMB, and OMB, that each contain a total mass (with the same generic albedo 
      and density assumptions) of
      6.2\,$\times$\,10$^{20}$,
      1.3\,$\times$\,10$^{21}$, and
      7.1\,$\times$\,10$^{20}$\,kg, 
      with a total of 26\,$\times$\,10$^{20}$\,kg.
      Therefore, although hundreds of thousands of asteroids 
      will still be discovered and they will provide valuable information about asteroids
      at small size scales, their expected contribution in terms of mass is minuscule 
      (below the part per million level). }

  \subsection{Applying the SDSS distribution to all asteroids \label{sec: SDSStoMPC}}
\begin{figure*}[t]
  \includegraphics[width=\textwidth]{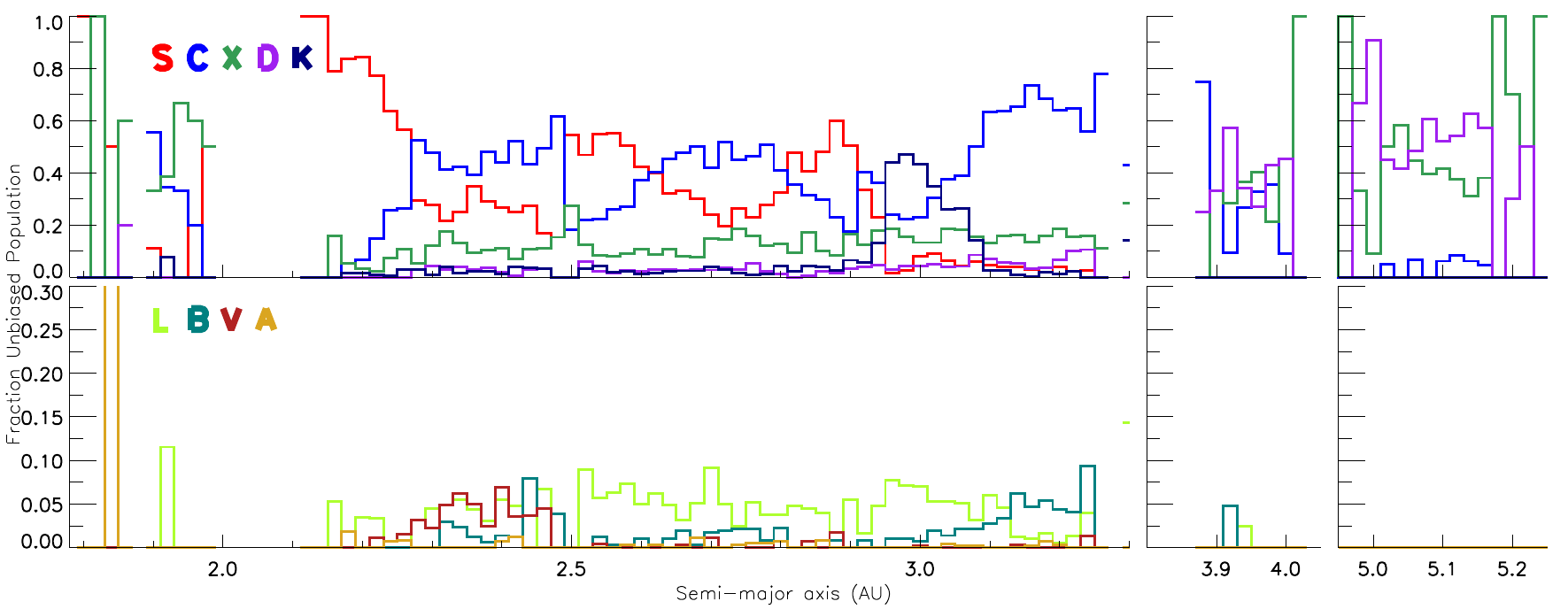}
  \caption[Debiased taxonomic distribution]{%
    The bias-corrected fraction of each class in each 0.02 AU bin according to
    SDSS data (each bin summing over all classes equals 100\%). All objects are 5
    km or larger. The distribution in this figure is dominated by smaller objects
    (H of 13 to 15.5). Because we are plotting by number of asteroids, the
    collisional families play an important role in this figure
    (\eg, the Vestoids in the inner belt). 
    \label{fig: classDEBIAS}
  }
\end{figure*}

\begin{figure*}[t]
  \includegraphics[width=\textwidth]{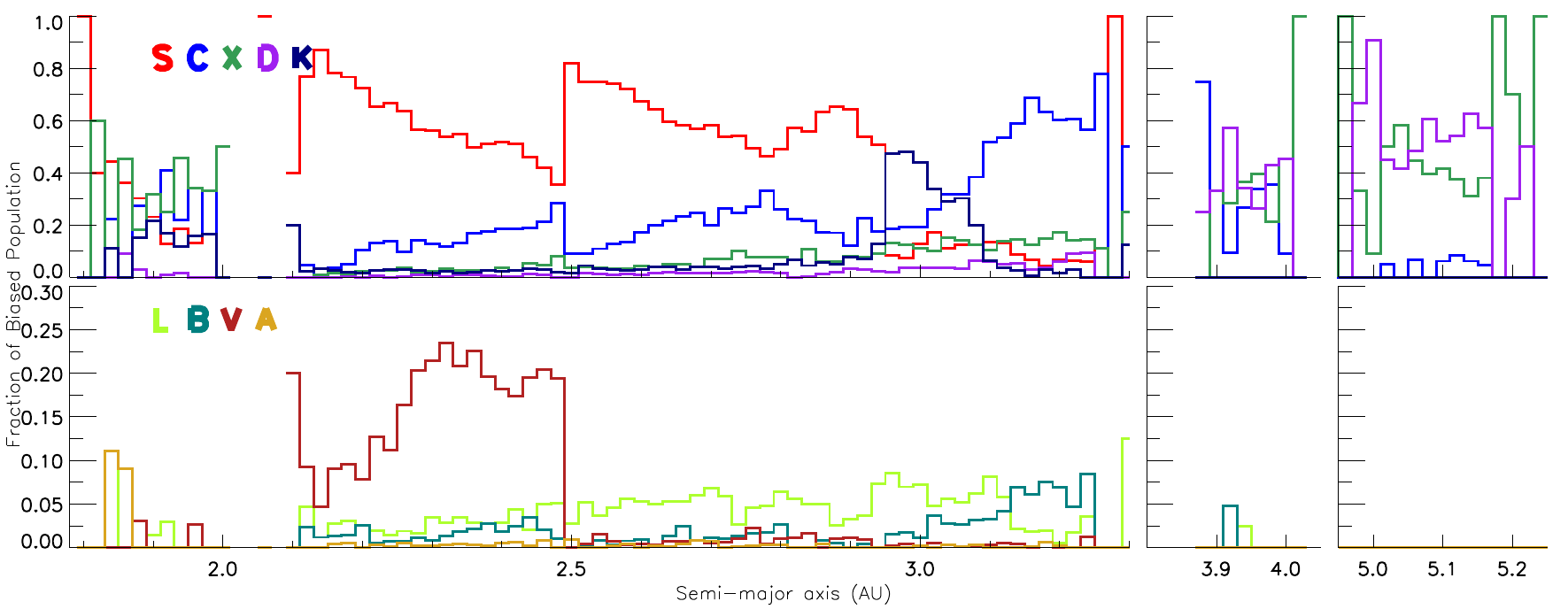}
  \caption[Biased taxonomic distribution]{%
    The observed fraction of each class in each bin according to SDSS data (each
    bin summing over all classes equals 100\%). In this case we did not cut
    the sample at a particular size range. The smaller, brighter S-types are
    more prevalent everywhere, and small, bright V-types make up nearly 20\%
    of the second half of the inner belt.  In this case we are plotting
    S-types smaller than 5 kilometers whereas we are not sampling the darker
    C-types of similar size. The difference between this figure and
    Fig.~\ref{fig: classDEBIAS}
    demonstrates the importance of correcting a sample for observational
    biases.  
    \label{fig: classBIAS}
  }
\end{figure*}

    \indent After applying data quality cutoffs, H-magnitude cutoffs, and 
    taxonomic classifications, we can now calculate the number of SDSS objects in each class
    according to size and distance. We use H magnitude bins of 1 magnitude
    ranging from 3 to 16 (though each class has its appropriate H magnitude
    cutoff listed in Table~\ref{tab: cuts}).
    The semi-major axis bins applied are 0.02 AU
    wide ranging from 1.78 to 5.40 AU. Only asteroids among Hungarias, the
    main belt, Cybeles, Hildas and Trojans are included in this study covering
    the distances 1.78-2.05, 2.05-3.27, 3.27-3.7, 3.7-4.2, and 5.05-5.40 AU,
    respectively. Near-earth objects, comets, Centaurs, Transneptunian objects
    and any other objects outside the mentioned zones were excluded. We
    calculate the number of objects in each bin \add{and} the fraction of each class
    in each bin (F$_i$, where i is the taxonomic class). For example, for objects
    with H magnitudes between 13 and 14 and semi-major axes between 2.30 and
    2.32 AU, we might find 60\% of the objects are S-type
    (F$_s$\,=\,0.6), 20\% are C-type
    (F$_c$\,=\,0.2), 20\% are X-type
    (F$_x$\,=\,0.2). 
    Figs~\ref{fig: classDEBIAS} and~\ref{fig: classBIAS}
    show the bias-corrected and biased fraction of
    objects. The biased view of the asteroid belt shows a dominance of S-types
    (by number) out to nearly 3 AU because of the inclusion of the abundant
    smaller, higher albedo bodies (whose small, dark counterparts, the C-types,
    were not observed). The bias-corrected version demonstrates that instead,
    the S-types and C-types alternate dominating by number throughout the
    belt. Asteroid families play an important role in these
    figures since they contribute large numbers of taxonomically similar
    objects.  

\begin{figure}[t]
  \includegraphics[width=.5\textwidth]{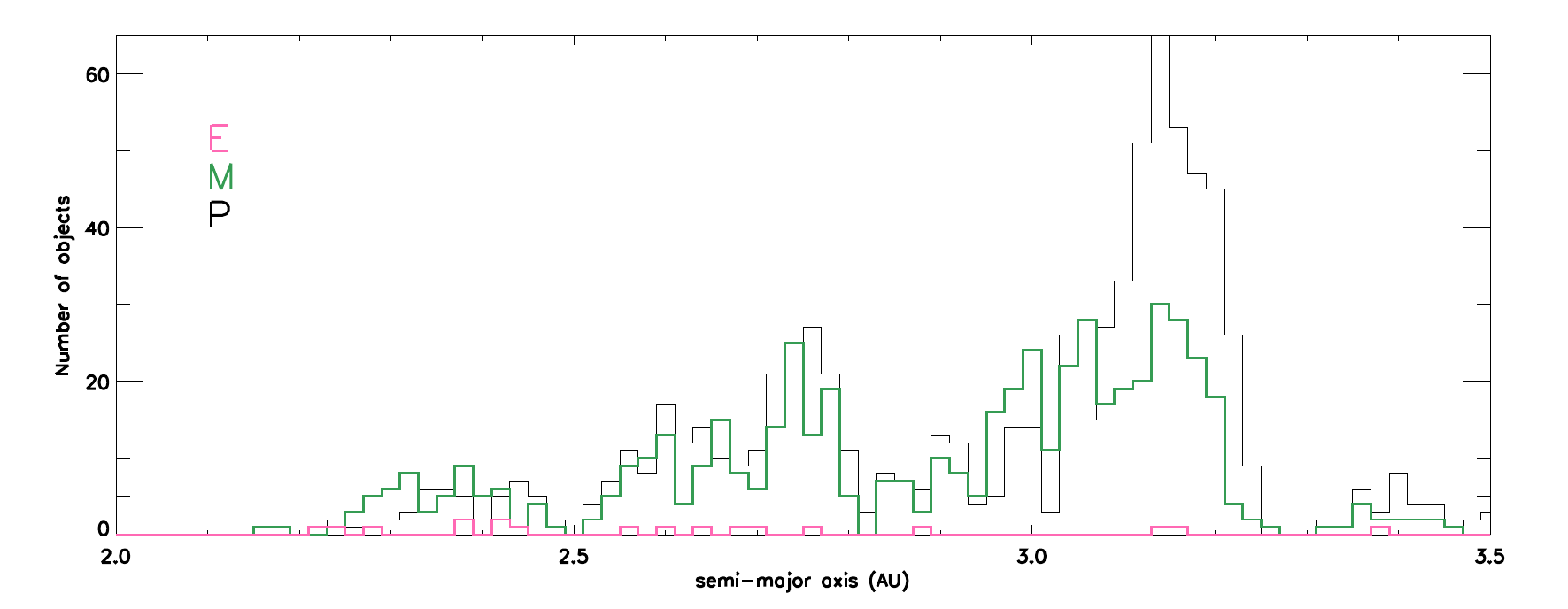}
  \caption[Biased distribution of EMP]{%
  The bias-corrected distribution of E, M, and P-type asteroids across the main
  belt based on the 1500 X-types in our sample (including spectral surveys and
  SDSS) with WISE, AKARI, or IRAS albedo determinations
  \citep{2010-AJ-140-Ryan, 2011-PASJ-63-Usui, 2011-ApJ-741-Masiero}.
  These objects are used
  to determine the relative fraction of M to P types among X class objects in
  each zone of the belt. Because only a subset of our SDSS X-types had albedos
  available we applied the fraction of M/P in each region to our entire X-type
  dataset. In the area near 3.0 AU we remove all SDSS objects classified as
  X-type in the Eos family. Because of the spectral similarity between X and
  K-types in SDSS colors, many K-types Eos family objects were classified as X
  (see Section~\ref{sec: mods} for discussion on classification ambiguity).
    \label{fig: empDEBIAS}
  }
\end{figure}

    \indent Albedo data enable the separation of X-types into three sub groups:
    E, M, P \citep{1984-PhD-Tholen}.
    Since albedo data are not available for every single spectral X-type,
    we calculate the fraction of E, M, and P for each region:
    Hungaria, Inner, Middle, Outer, Cybele, Hilda, Trojan. This fraction is
    calculated based on $\sim$2000 X-type objects in our sample with
    albedo measurements (from a total of 2500 X-types)
    from IRAS, AKARI, and WISE
    \citep{2010-AJ-140-Ryan, 2011-PASJ-63-Usui, 2011-ApJ-741-Masiero}. 
    See Fig.~\ref{fig: empDEBIAS}
    for the bias-corrected distribution of
    the E, M, and P types across the main belt that was used to extrapolate
    the X-type EMP fraction for our entire dataset. Among Hungarias the sample
    is entirely E-type as expected. There are an insignificant number of
    E-types among the other regions (though we note a bias against observing
    high visible albedo objects in mid-infrared wavelength ranges). The
    fraction of all bias-corrected X-types that are M in each region are: 0.00,
    0.58, 0.44, 0.35, 0.28, 0.08, and 0.17, respectively. The fraction for
    P-types is thus one minus the M-type fraction, except for the Hungaria
    region where it is also zero. Among Trojans we find that \add{0.17 (1 out
      of 6)} X-types have 
    an M-type albedo, however because of large uncertainty due to a small sample
    we assume the same fraction for Trojans as Hildas (0.08).  

    \indent We now know the relative abundance of each taxonomic type at each size range and distance
    determined from the SDSS dataset with and adjustment for the division of
    E, M, and P types from the X class.  
    However, at many size ranges the SDSS only observed $\sim$30\%
    of the total asteroids that exist at that size and distance. As long as we only use a 
    size range in which asteroid discovery is essentially complete or make a correction for 
    discovery incompleteness, we can apply these fractions to
    the entire set of known asteroids at these sizes from the Minor Planet Center to determine 
    the distribution of taxonomic type across the main belt according to number, surface area, volume,
    and mass. \\

     When calculating the number of objects, surface area, volume, or mass at
     each size range and distance  
     we use two different methods. For the largest asteroids with H\,$<$\,10 where our SDSS sample 
     is complete, we calculate the surface area, volume, or mass for each asteroid using
     that body's H magnitude, albedo (or average albedo for its taxonomic class when not available),
     and average density \citep{2012-PSS-73-Carry} for that taxonomic class.
     
     For objects with 10\,$<$\,H\,$<$\,13, where our sampling is not complete, we use the 
     following method.  The surface area, volume, or
     mass is calculated for an object using the H magnitude, average albedo and average 
     density for that class. That value is multiplied by the number of objects
     of that class in that bin (N$_i$) which is the the total number 
     of known asteroids in that (size and distance) bin, N$_{bin}$, and 
     by the fraction (F$_i$) of objects of that class from SDSS:
     N$_i$=N$_{bin}$\,$\times$\,F$_i$. 
  
     For objects with H$>$13 we have the added complication that we cannot directly apply 
     \add{our fraction to the total number of known objects because} our
     fraction of each type at each  
     size from SDSS is calculated with certain (higher albedo) classes removed.
     We thus must also calculate the fraction of objects in the SDSS database that 
     were kept, F$_{kept}$, (i.e., those that were not removed because they are smaller 
     than $\sim$5 kilometers) for H magnitude bins H=13, 14, and 15. For all other size 
     bins F$_{kept}$ is equal to 1. The number of objects of a certain class (N$_i$) can 
     be determined by the total number of discovered objects in that bin (N$_{bin}$) multiplied 
     by the fraction of objects in that class (F$_i$) and by the fraction of objects in that bin 
     that are kept (F$_{kept}$), thus
     N$_i$=N$_{bin}$\,$\times$\,F$_i$\,$\times$\,F$_{kept}$.

     Previously in this section, we calculated the bias-corrected fraction of 
     E, M, and P types in each zone, although, as above, we cannot apply 
     this true (bias-corrected) correction factor to the observed (biased) MPC 
     dataset. For the H bins 14 and 15 where some M-types were removed due 
     to size we calculate the fraction of (M+P)-types kept in each region. 
     The fraction for H=14 is 0.60, 0.67, 0.64, 0.79, for the IMB, MMB, OMB, 
     and Cybeles and 0.22, 0.39, 0.41, 0.70 for H=15. There are no small 
     objects in our sample to be removed in the Hilda and Trojan regions so the fraction kept is unity.

     Finally, if we simply assume an average H magnitude for each bin 
     (say 12.5 for the H=12 bin) we could potentially over- or underestimate the 
     surface area, volume, or mass, depending on the H magnitude distribution 
     of objects in that bin. We thus calculate the number of objects in each 0.1 H 
     magnitude sub-bin and apply the same class fractions to each for accuracy.

\section{The compositional makeup of the main belt \label{sec: distrib}}

  \subsection{Motivation for number, surface area, volume, and mass \label{sec: justify}} 

    \indent Previous work calculated compositional distribution based on the number of
    objects at each distance
    \citep[e.g.][]{1975-Icarus-25-Chapman, 1982-Science-216-Gradie,
      1989-AsteroidsII-Gradie, 2003-Icarus-162-Mothe-Diniz}.
    This was not unreasonable because those datasets
    included only the largest objects, often greater than 50\,km in diameter. \\
    \indent If we restrict our study to the number of asteroids, our views
    would be strongly influenced by the small asteroids.      
    There are indeed more asteroids of small size than large ones. This is the 
    result of eons of collisions, grinding the asteroids down from 
    larger to smaller. 
    The size-frequency distribution of asteroids
    (Fig.~\ref{fig: sdsscomplete}) can be approximated by a power-law, and
    for any diameter below 20\,km, there are about 10 times more asteroids
    with half the diameter.
    The amount of material (\ie, the volume) of the two size ranges is
    however similar: if there are $n$ asteroids of a given diameter $D$,
    there are about 10$n$ asteroids with a diameter of $D/2$, each with a
    volume 8 times smaller, evening out the apparently dominating importance
    of the smaller sizes. 
    \add{Ceres alone contains about a third of the mass in the entire main belt using
      a mass of 30\,$\times$\,10$^{20}$\,kg for the main
        belt \citep{2013-Icarus-222-Kuchynka}, and 9\,$\times$\,10$^{20}$ kg for Ceres
        \citep[from a selection of 28 estimates, see][]{2012-PSS-73-Carry}),
        and yet it is negligible (1 out of 600,000) when 
      accounted for in a distribution by number.}
    Therefore, the relative importance of Ceres in the main belt can change by
    orders of magnitude depending on how we look at the distribution.\\
    \indent The study of the compositional distribution by
    number is perfectly valid and is useful for size-frequency distribution studies
    and collisional evolution. For studies of the distribution of the amount
    of material, it puts too much emphasis on the 
    small objects compared to the largest.
    A simple way to balance the situation is to consider each object weighted
    by its diameter. 
    This opens new views on asteroids: we can study how much surface area
    of a given composition is accessible for sampling or mining purposes
    (Section~\ref{sec: surface}), or how much material accreted
    in the early solar system has survived in the Belt 
    (Sections~\ref{sec: volume} and~\ref{sec: mass} for the distributions
    by volume and  mass).

  \subsection{Asteroid distribution by surface area\label{sec: surface}}

\begin{figure*}[t]
  \includegraphics[width=\textwidth]{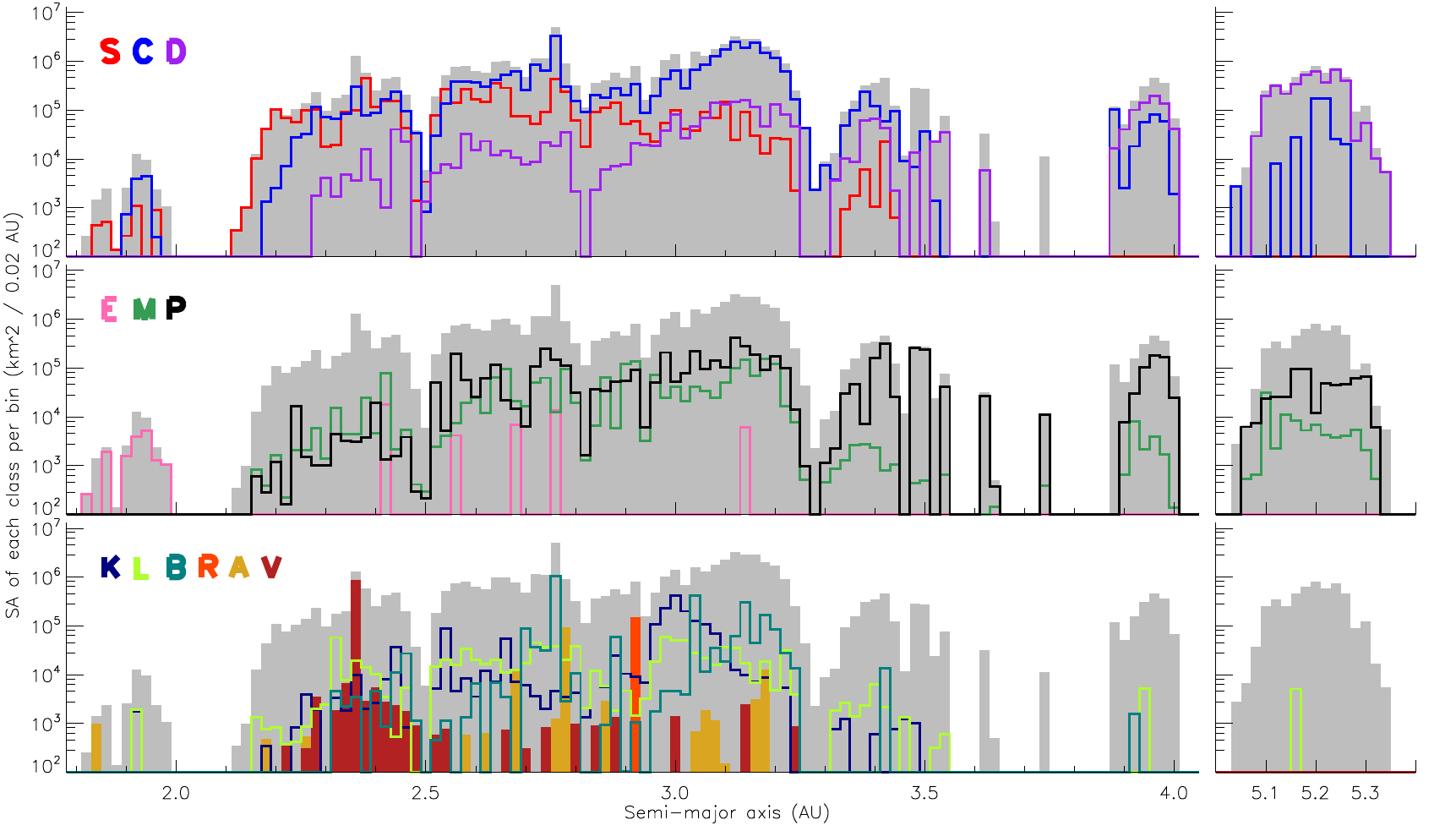}
  \caption[Debiased distribution, by surface area]{%
        The surface area  (km$^2$) of each taxonomic class in each 0.02 AU. 
        The y-axis scale is logarithmic to include all classes on the same scale. 
        All objects are 5 km or larger. While we don't classify R-types in our SDSS dataset,
    the one R-type in this plot is (349) Dembowska, from the spectroscopic surveys.
    \label{fig: classSurface}
  }
\end{figure*}

    \indent To estimate the surface area of each asteroid, we need first to
    estimate its diameter $D$. For that we use the following equation
    from \citet[][and references therein]{2007-Icarus-190-Pravec}:

\begin{equation}
  D = \frac{1329}{\sqrt{a}} 10^{-\ 0.2\ H} \label{eq: diam}
\end{equation}

    \noindent where $H$ is the absolute magnitude (determined by the 
    SDSS survey) and $a$ is the albedo.
    For the largest asteroids (H\,$<$\,10) we use the object's calculated albedo 
    from WISE, AKARI, or IRAS. For small asteroids, and large ones for
    which no albedo is available, we use the average albedo for that object's
    taxonomic class, see \ref{sec: albedo}. 
    The equation above provides a crude estimation of the diameter only.
    Evaluation for a particular target should be considered with caution,
    the absolute magnitude and albedo being possibly subject to large
    uncertainties and biases \add{ \citep[e.g.,][]{2005-Icarus-179-Romanishin,2011-AJ-141-Mueller,2012-Icarus-221-Pravec}.}
    We can nevertheless make good use of this formula for statistical
    purposes: the precision on the diameters is indeed rough but seemingly
    unbiased \citep{2012-PSS-73-Carry}.
    With a diameter $D$ determined for each asteroid, we estimated their
    individual surface area $\mathcal{S}$
    by computing the area of a sphere of the same diameter: 
     $\mathcal{S}$\,=\,$\pi D^2$.
    The surface area distribution is presented in
    Fig.~\ref{fig: classSurface}.\\
    \indent The total surface area per bin ranges from 10$^3$\,km$^2$ in the
    Hungarias to 10$^6$ in the main belt.
    Viewing the distribution with respect to surface area we can
    immediately notice the relative importance of larger bodies. Ceres 
    and Pallas are represented by the blue peak near 2.75\,AU and 
    Vesta by the red peak near 2.35\,AU. Additionally the E-types in pink,
    distributed throughout the main belt, are each only one or two asteroids.

  \subsection{Asteroid distribution by volume\label{sec: volume}}

    While the real value we seek is mass, because the density 
    contributes significant uncertainty to the mass calculation we also present the
    distribution according to volume of material which gives similar results but is not affected by 
    density uncertainties. 
    
    \indent To evaluate the amount of material in the main belt, for each taxonomic
    class, we estimate the volume $\mathcal{V}$ of all the asteroids by
    computing the volume of a sphere of the same diameter:
    $\mathcal{V} = \pi D^3 / 6$.
    We use the same method to calculate the total volume distribution
    by applying SDSS taxonomic fractions to the MPC dataset
    as described in Section \ref{sec: SDSStoMPC}.
    By looking at the compositional distribution in terms of volume instead of
    numbers, most of the issues described in Section~\ref{sec: justify} are
    addressed.
    Indeed, if there are about 2500 asteroids with a diameter of 10 km in
    the main belt, their cumulated volume is 300 times smaller than that of
    Ceres, re-establishing the proportions.
    The conversion from numbers to volume also corrects our sample for an
    overrepresentation of the contribution by collisional families (when viewed by number). 
    Indeed, a swarm of fragments is released
    during every cataclysmic disruptive event, ``artificially'' increasing the
    relative proportion of a given taxonomic class locally (\eg, the Vestoids
    in the inner belt, see Fig.~\ref{fig: classBIAS}). Here, 
    we are accounting for all the material of the family as if put back
    together again.

\begin{figure*}[t]
  \includegraphics[width=\textwidth]{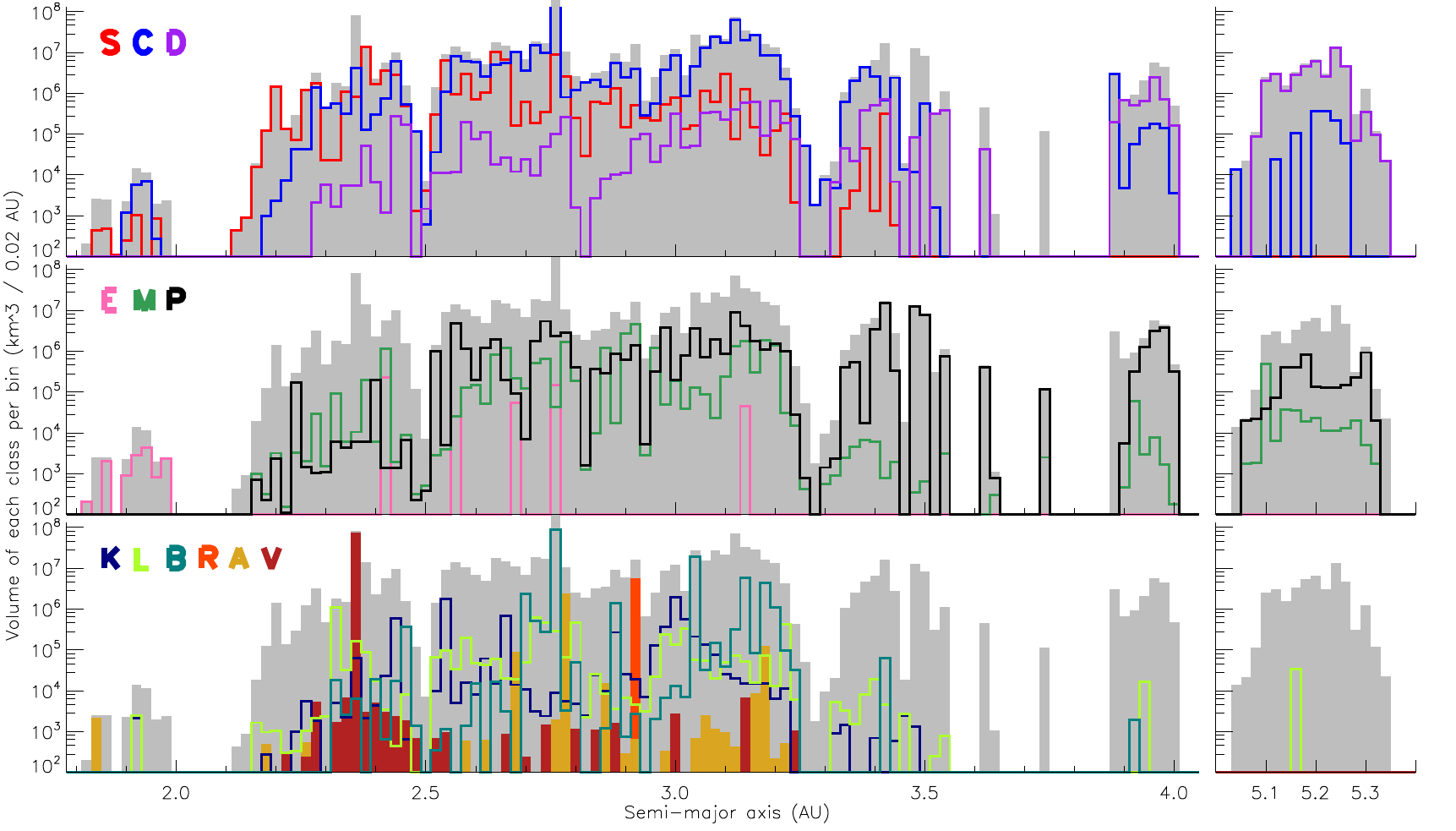}
  \caption[Debiased distribution, by volume]{%
    The volume  (km$^3$) of each taxonomic class in each 0.02 AU. The y-axis scale 
    is logarithmic to include all classes on the same scale. All objects are 5 km or larger. 
    \label{fig: classVolume}
  }
\end{figure*}

    We present the distribution of taxonomic class by volume in
    Fig.~\ref{fig: classVolume}. The distribution is the same as for
    surface area, but with the y-axis stretched because volume 
    is proportional to diameter cubed while surface area is 
    proportional to diameter squared. Asteroid distributions by
    volume were first presented by \citet{2012-DPS-Consolmagno}.

  \subsection{Asteroid distribution by mass\label{sec: mass}}
    \indent Ultimately, the mass is the
    physical parameter we seek that provides insights on the
    distribution of material in the solar system.
    To precisely measure the mass of each asteroid we would need a fleet of missions
    to fly by each asteroid.
    Barring that as an option in the foreseeable future, 
    to \textit{estimate} the mass of each asteroid we need an approximate density 
    together with the estimated volume determined above.
    The density is the least well-constrained value used in this work because
    these measurements are extremely difficult to obtain
    \citep[see discussion in][]{
      2002-AsteroidsIII-4.2-Britt, 2006-LPI-37-Britt, 2012-PSS-73-Carry}.
    Nevertheless, the study of meteorites tells us that the available range
    for asteroid density is narrower than it may seem.
    Indeed, no meteorite denser than 7.7 g/cm$^3$~has ever been found, and 
    most of the meteorites cluster in a tight range, from 2 to 5
    g/cm$^3$~\citep[see][and references therein]{1998-MPS-33-Consolmagno,
      2008-ChEG-68-Consolmagno, 2003-MPS-38-Britt, 
    2010-MPS-45-Macke, 2011-MPS-46-Macke}, with the exception of iron
    meteorites above 7 g/cm$^3$~\citep[see the summary table
      in][]{2012-PSS-73-Carry}.
    This range may be wider, especially at the lower end, for
    asteroids due to the possible presence of voids in their
    interiors, such as the low density of 1.3 g/cm$^3$~found for 
    asteroid (253) Mathilde \citep{1997-Science-278-Veverka}.
    However, even if we assign an incorrect density to an asteroid, the impact
    on its mass will remain contained within a factor of 4 at the very worst.
    The impact may even be smaller as the typical densities of the most common
    asteroid  classes (\ie, C and S) are known with better accuracy
    \citep{2012-PSS-73-Carry}.
 
    \indent The uncertainty on the density will therefore affect the
    distribution in a much lesser extent than equal weighting of bodies according to number.
    Of course, any uncertainty on any of the parameters will sum up in the
    total uncertainty. However, we are confident that the trends we discuss
    below are real: both the discovery completeness, diameter estimates, and
    average albedo and density per taxonomic class have become more and more
    numerous and reliable over the last decade.
    
    To calculate the distribution by mass we apply the average density of each 
    class (Table~\ref{tab: cuts},~\ref{sec: density}) and multiply that by the 
    volume determined in the previous Section.
      For Ceres, Vesta, Pallas, and Hygiea, 
    the four most massive asteroids we include their measured 
    masses 
    \citep[9.44, 2.59, 2.04, and 0.86\,$\times$\,10$^{20}$\,kg,
      from][]{2012-PSS-73-Carry, 2012-Science-336-Russell}  
    each accounting for about 31\%, 9\%, 7\% and 3\% of the mass of the 
    main belt, respectively
    \citep[using a total mass of the belt of
      30\,$\times$\,10$^{20}$\,kg,][]{2013-Icarus-222-Kuchynka}

\begin{figure*}[t]
  \includegraphics[width=\textwidth]{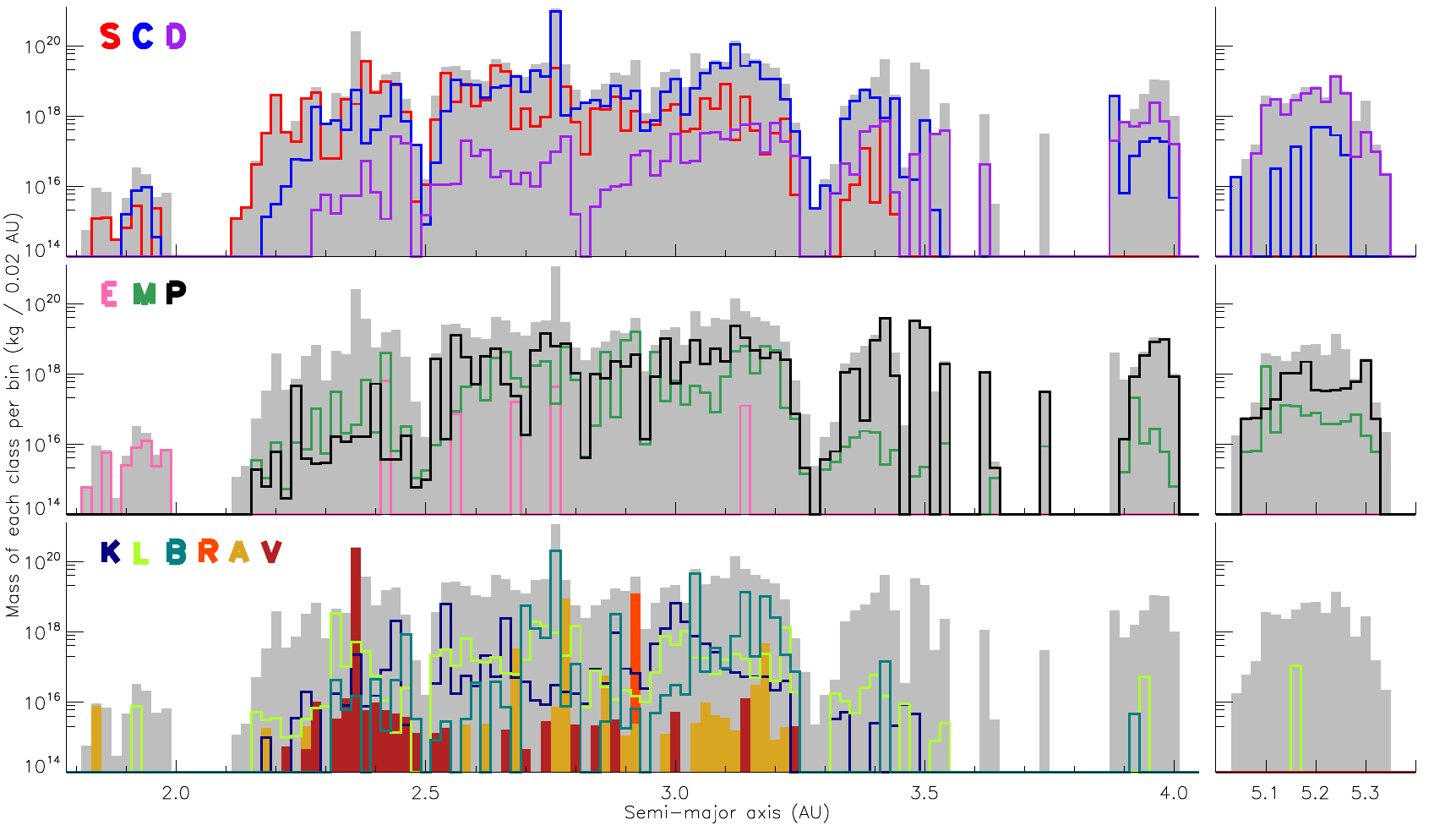}
  \caption[Debiased distribution, by mass]{%
    The mass (kg) of each taxonomic class in each 0.02 AU bin.
    All objects are 5 km or larger.
    \label{fig: classMass}
  }
\end{figure*}

\begin{figure}[t]
  \includegraphics[width=.5\textwidth]{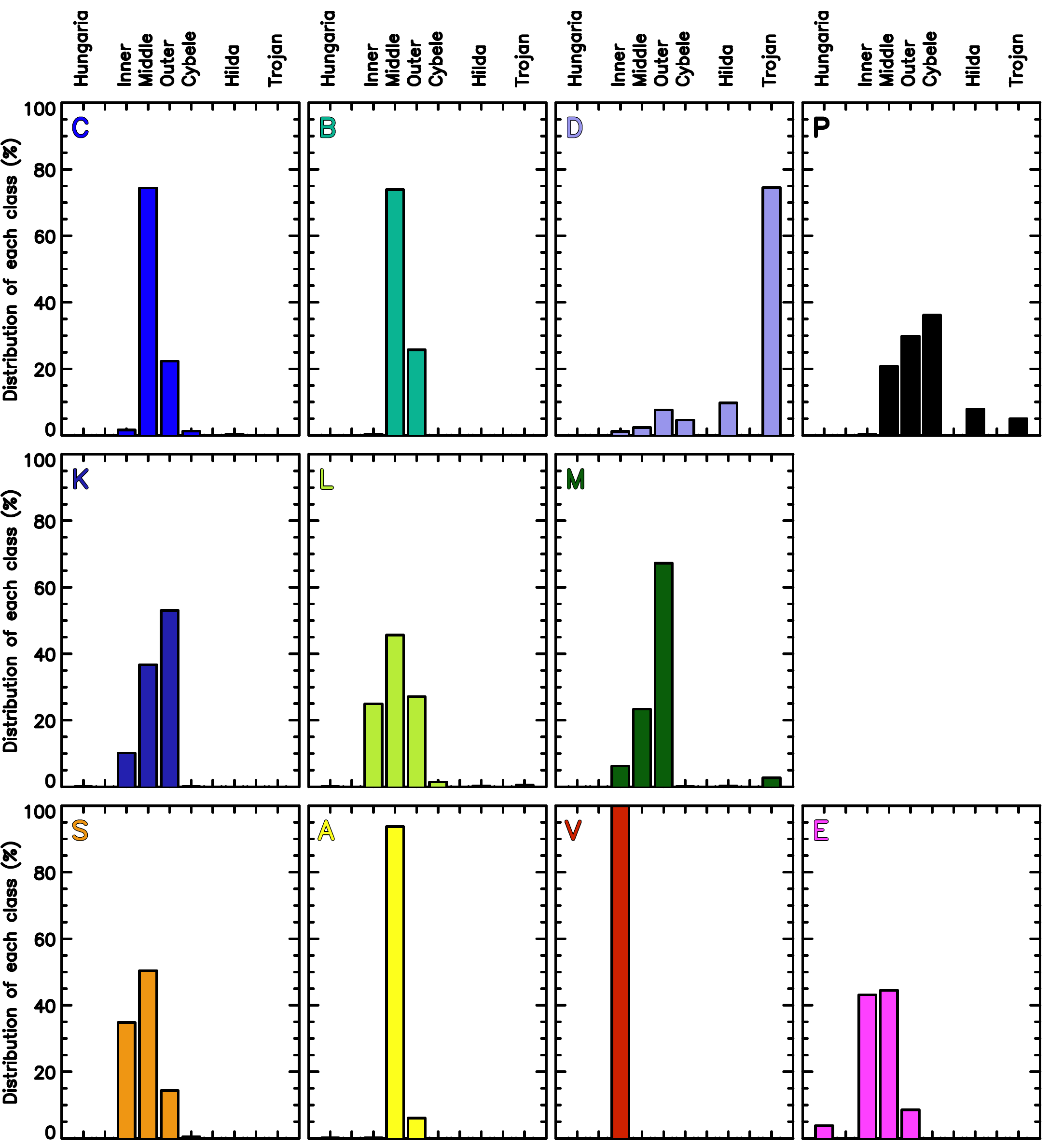}
  \caption[Fraction of class per region]{%
   The fractional mass distribution of each class across the belt.
   The total of each class across all zones sums to 100\%.
    \label{fig: bars}
  }
\end{figure}

      \indent The distribution of mass is presented in Fig.\ref{fig: classMass}
      The fractional distribution of each class throughout the belt
      is given in Fig.~\ref{fig: bars}.     
       Again, we find the general trends to be similar to volume and surface area.
      The difference in the case of mass is that the relative abundance of 
      the taxonomic types have changed. Because S-types are generally
      denser than C-types by a factor of roughly 2 we see S-type material
      is more abundant relative to C than in our previous plots. Because 
      in many cases the relative abundance of different taxonomic types
      already vary by an order of magnitude or more, we do not see 
      drastic relative abundance changes. For example, C-types contribute
      more mass to the outer belt than S-types even though their relative
      abundance by mass is closer than by volume.

    \subsection{Search for S-types among Hildas and Trojans}

      \indent We note a sharp cliff at the edge of the outer belt delineating 
      the limit of S-type asteroids. \citet{2003-Icarus-162-Mothe-Diniz}
      were the first to show the presence of S-types out to 3\,AU in
      their dataset of asteroids 15\,km and greater. We find 
      that almost no S-types exist among Cybeles, and they are entirely
      absent beyond 3.5\,AU. 

      Despite the bias toward discovering S-types 
      (they reflect 5 times more light than C-, D-, and P- type bodies of the same size)
      and their abundance in the main belt, 
      we find no convincing evidence for S-type asteroids among Hildas and Trojans.
      Ten asteroids among Hildas and Trojans have at least one SDSS
      measurement classified as S-type. Half of those objects have another
      observation that does not suggest an S-like composition (the second
      observation is typically classified D). Visual inspection 
      suggests the quality of the data for two of them are poor.
      Only one object has an albedo measurement, but the low
      value of 0.07 is very unlikely to represent an S-type composition. One object among
      each of the Hildas and Trojans remains. While we cannot rule out these
      objects, given the other mis-categorized data and our caution against
      interpreting single objects, we do not find any convincing evidence from
      this dataset of S-types among Hildas or Trojans
      (or any other high albedo classes).  We reach conclusions similar 
      to the many authors who have investigated the compositions of these regions 
      \citep{2003-Icarus-164-Emery, 2004-Icarus-170-Emery, 2011-AJ-141-Emery, 
        2004-Icarus-172-Fornasier, 2007-Icarus-190-Fornasier, 
        2007-AJ-134-Yang, 2011-AJ-141-Yang, 
        2008-AA-483-Roig, 2008-Icarus-193-Gil-Hutton, 
        2011-ApJ-742-Grav, 2012-ApJ-744-Grav, 
        2012-ApJ-759-Grav}. The wider range of albedos found among the smallest
        Trojans \citep{2009-AJ-138-Fernandez,2012-ApJ-759-Grav} which are not well-sampled in this work
        should prompt further follow up investigation of these targets
        to determine their taxonomic class. While it is possible this albedo 
        difference with size is due to the younger age of the smaller bodies 
        \citep{2009-AJ-138-Fernandez}, finding a wider variety of classes
        would prove interesting in the context of current dynamical theories
        such as by \citet{2005-Nature-435-Morbidelli}.

    \subsection{Evidence for D-types in the inner belt}

      \indent 
      We find evidence for D-types in the inner
      and mid belt from SDSS colors. The potential presence of 
      D-types was also seen by \citet{2010-AA-510-Carvano}. Here
      we take a scrutinizing look at the SDSS data to be certain the
      data are reliable.  
      
      While D-types typically have a low albedo,
      Bus-DeMeo D-types have been measured to have albedos as high as 0.12 (Bus
      and Tholen D-types have maximum albedos of 0.25). We
      compare the median albedo of D-types in the inner, middle, and outer
      belt. For samples of 35, 81 and 108 we find medians of 0.13, 0.13, and
      0.08. The median albedos in the inner and middle belt suggest that there is
      more contamination from other asteroid classes, however, there is still a
      large portion of the sample with low albedos. Next we inspect the data for
      all SDSS D-types in the inner belt including those without albedos. 
      We find that 9 out of the 65 objects
      were observed more than once and that they all remain consistent with a D
      classification, all objects observed twice were twice classified as D,
      objects with more observations were classified as D for at least half the
      observations. Additionally, we check if any of the 65 D-types are members
      of families. We find two objects associated with the Nysa-Polana
      family. Because there are many C- and X-types in that family it could
      indicate those two objects were misclassified, however, they represent a
      small fraction of our sample. Because many these objects have low albedos,
      are not associated with C- or X-type families and have been observed
      multiple times and remain consistent with the D class, we have confidence
      in the existence of D-types in the inner belt.  The orbital elements of
      inner belt D-types are scattered; we find no clustering of objects. The
      presence of D-type asteroids in the inner belt might not be entirely consistent with the
      influx of primitive material from migration in the Nice model.
      \citet{2009-Nature-460-Levison} find that D-type and P-type material do not come closer
      than 2.6 AU in their model, however, their work focused on bodies with diameters
      greater than 40\,km.

\section{Overall View \label{sec: discuss}}
  \add{We find a total mass of the main belt of 2.7\,$\times$\,10$^{21}$ kg
    which is in excellent agreement with the estimate by
    \citet{2013-Icarus-222-Kuchynka} of 
    3.0\,$\times$\,10$^{21}$ kg. 
  The main belt's most massive classes are C, B, P, V and S in 
  decreasing order (all B-types come from the spectroscopic sample, not the SDSS sample,
  see Sec.~\ref{sec: mods}). 
  The total mass of each taxonomic class and respective percentage of the total main 
  belt mass is listed in Table~\ref{tab: masstot}. 
  The overall mass distribution is heavily skewed 
  by the four most massive asteroids, (1) Ceres, (2) Pallas, (4) Vesta 
  and (10) Hygiea, together accounting for more than half of the mass 
  of the entire main belt. Ceres, Pallas, Vesta, Hygiea are roughly 35\%, 10\%, 8\%, and 3\% 
  respectively of the mass of the main belt (based on the total mass from this work). 
  If we remove the four most massive bodies as shown in Table~\ref{tab: masstot}, 
  the most massive classes are then C, P, S, B and M in decreasing order. 
  The mass of the C class is six times the mass of the S class, 
  and with Ceres and Hygeia removed, the S-types are about 1/3 and C-types
  2/3 of their combined mass. } \\
%
%
%
%
\begin{table*}[ht]
\begin{center}
\begin{tabular}{cr@{\,$\times$\,}lrr}									
\hline
\hline
Class & \multicolumn{2}{c}{Mass	(kg)} & Fraction (\%) & Largest Removed	(\%)\\			
\hline									
A	& 9.93 & $10^{18}$ &       0.37	&  0.37	\\
{\bf B}	& 3.00 & $10^{20}$ & {\bf 11.10} & {\bf  3.55} 	\\
{\bf C}	& 1.42 & $10^{21}$ & {\bf 52.53} & {\bf 14.41}	\\
D	& 5.50 & $10^{19}$ &       2.03	&  2.03	\\
K	& 2.56 & $10^{19}$ &       0.95	&  0.95	\\
L	& 1.83 & $10^{19}$ &       0.68	&  0.68	\\
{\bf S}	& 2.27 & $10^{20}$ & {\bf  8.41} &  {\bf 8.41}	\\
{\bf V}	& 2.59 & $10^{20}$ & {\bf  9.59} &  {\bf 0.01}	\\
E	& 1.46 & $10^{18}$ &       0.05	&  0.05	\\
M	& 8.82 & $10^{19}$ &       3.26	&  3.26	\\
{\bf P}	& 2.98 & $10^{20}$ & {\bf 11.02} & {\bf 11.02}	\\
\hline
Total	& 2.70 & $10^{21}$ &     100	&	45	\\
\hline
\end{tabular}									
  \caption[Mass totals for each Taxonomic Type]{%
    Total mass of each taxonomic type. We present the total mass and
    fractional mass of each type. The last column is the percentage with the
    four most massive asteroids (Ceres, Pallas, Vesta, and Hygiea)
    removed. While we list the values two 2 decimal places as the mathematical
    result we do not claim accuracy to that level. 
       \label{tab: masstot}
  }
\end{center}
\end{table*}
%
%
%
%
    \indent \add{The distribution of each class by total mass percentage in each 
    zone of the main belt is shown in Table~\ref{tab: masszone}. As we 
    expect, E-types dominate the Hungaria region both by mass percentage 
    and also in total number of objects, and C and S-types are the next most 
    abundant by mass in the Hungaria region.
    Most of the mass of the inner belt is in Vesta, and S-types account for 4 times
    more mass in the inner belt than C-types ($\sim$20 and $\sim$5\% of the 
    total mass, respectively).
    In the middle belt Ceres and Pallas once again make up the majority of the mass. When excluding
    these two bodies, C-types and S-types each make up $\sim$30\% of the mass
    of the middle belt, P-types $\sim$20\% and B- and M-types $\sim$5\%.
    The outer belt is heavily weighted toward C-types including or excluding the
    most massive body, (10) Hygiea.
    A shift to an abundance of P-types occurs in the Cybeles.
    Both the Cybeles and Hildas are predominantly P-type by mass. The majority of Trojans 
    are D-type asteroids. Based on these findings, we can confirm and recreate the
    general trend of
    E, S, C, P, and D-type asteroids with increasing distance from the sun as 
    established by \citet{1982-Science-216-Gradie}
    and \citet{1989-AsteroidsII-Gradie}.} \\
%
%
%
%
%
\begin{table*}[ht]
\begin{center}
\begin{tabular}{lrrrrrrrrrrrr}
\hline
\hline
Zone       &  A  &   B   &   C   &   D   &   K  &     L &     S &   V   &     E &  M   &   P   & Total \\
\hline
Hungaria	&	7	&	0	&	21	&	1	&	5	&	7	&	9	&	0	&	50	&	0	&	0	&	100	\\
Inner	&	0	&	$<$1	&	6	&	$<$1	&	1	&	1	&	21	&	69	&	$<$1	&	1	&	$<$1	&	100	\\
Middle	&	$<$1	&	15	&	70	&	$<$1	&	1	&	1	&	8	&	0	&	$<$1	&	1	&	4	&	100	\\
Outer	&	$<$1	&	13	&	52	&	1	&	2	&	1	&	5	&	0	&	$<$1	&	10	&	15	&	100	\\
Cybele	&	0	&	$<$1	&	13	&	2	&	$<$1	&	$<$1	&	1	&	0	&	0	&	$<$1	&	84	&	100	\\
Hilda	&	0	&	0	&	14	&	15	&	0	&	$<$1	&	0	&	0	&	0	&	1	&	71	&	100	\\
Trojan	&	0	&	0	&	2	&	67	&	0	&	$<$1	&	0	&	0	&	0	&	4	&	26	&	100	\\
\hline
\end{tabular}
  \caption[Percent of mass per zone]{%
    Percentage of mass distributed through each each zone. The total
    for each zone summed over all classes equals 100\%. \add{In some zones there were 
    very few ($<$20) objects of a certain class. We note these here to be aware of possible 
    uncertainties do to small number statistics: A-types in all zones, B-types in Hildas, C-types
    in Trojans, D-, K-, L-, and S-types in Cybeles, V-types in the middle and outer belt, and X-types 
    in Trojans.}
    \label{tab: masszone}
  }
\end{center}
\end{table*}
%
%
%
%
\begin{table*}[ht]
\begin{center}
\begin{tabular}{lrrrrrrrrrrr}
\hline
\hline
Zone     &  A   & B     &  C     &  D     &   K     &  L     & S     & V     &E     & M     & P      \\
\hline
Hungaria	&	$<$1	&	0	&	0	&	0	&	$<$1	&	$<$1	&	0	&	0	&	3	&	0	&	0	\\
Inner	&	$<$1	&	$<$1	&	1	&	1	&	10	&	25	&	35	&	100	&	44	&	6	&	$<$1	\\
Middle	&	94	&	74	&	74	&	2	&	37	&	47	&	51	&	0	&	45	&	23	&	21	\\
Outer	&	6	&	26	&	22	&	8	&	53	&	27	&	14	&	0	&	8	&	67	&	30	\\
Cybele	&	0	&	$<$1	&	1	&	5	&	$<$1	&	1	&	$<$1	&	0	&	0	&	$<$1	&	36	\\
Hilda	&	0	&	0	&	$<$1	&	10	&	0	&	$<$1	&	0	&	0	&	0	&	$<$1	&	8	\\
Trojan	&	0	&	0	&	$<$1	&	74	&	0	&	$<$1	&	0	&	0	&	0	&	3	&	5	\\
\hline
Total    & 100& 100 &  100 &  100 &   100 &  100 & 100 & 100 & 100 & 100 & 100 \\
\hline
\end{tabular}
  \caption[Percent of mass per class]{%
    Percentage of mass distribution for each class in each zone. The percentage for each class
    summed over all zones equals 100\%. 
       \label{tab: massclass}
  }
\end{center}
\end{table*}
%
 %
  %
  \indent Previous work reports the distribution as relative frequency in each 
  semi-major axis bin. Thus the results reported so far can be 
  compared to previous work, however, previous fractions were reported by
  number and not by mass. This view ignores the relative importance of 
  each semi-major axis zone. The results
  must be weighted according to how much mass each region holds. The Hungarias
  only contain about 0.02\% of the mass of the inner belt for example. The 
  total mass of each region increases moving outward, peaks in the middle
  belt and decreases thereafter, though the Trojans hold more mass than
  the Hildas. Excluding the 4 most massive objects the total mass peaks
  in the outer belt. 
  
  A relative weighting by mass allows us to more 
  accurately see how each class is distributed across the belt (see Table~\ref{tab: massclass}). 
  For example there have been differing views about S-type abundances.
  S-types are typically thought of being most abundant 
  in the inner belt where their relative frequency is greatest 
  \citep{1989-AsteroidsII-Gradie,
    2002-Icarus-158-BusII}. \citet{2003-Icarus-162-Mothe-Diniz} find S-types
  distributed evenly across the main belt.
  We report that one third of the mass of all S-types is in the inner belt,
  one half is actually in the middle,
  belt and $\sim$15\% of S-type mass is in the outer belt.
  
  Another example of the importance of relative weighting is the E-types.
  E-types are typically associated with the Hungaria region. The
  numerous E-types \citep{1984-PhD-Tholen,2004-AJ-128-Clark}
  found in this region are thought to be a part of the Hungaria
  asteroid family \citep{2009-Icarus-204-Warner,2010-Icarus-207-Milani},
  the largest being (434) Hungaria after which this region is named.
  However, Hungarias only account for 3\% of the mass of E-types. 
  Nearly 90\% of the mass of E-types resides in the inner and 
  middle belt split among a few large asteroids. 	 
  
  While the large majority of V-type mass is contained within Vesta
  (although one must keep in mind this is a differentiated body
  and thus has differing composition as a function of distance 
  from the center, only the surface layer is V-type), aside from Vesta nearly 20\% of the mass of 
  V-types is in the outer belt, due to (1459) Magnya. A careful
  study of the distribution of V-types across the belt with follow-up
  observations of SDSS candidates was performed by \citep{2008-Icarus-198-Moskovitz}. 
  
  Most of the mass of C- and B-types lies in the middle belt, however,
  if Ceres and Pallas are excluded, the majority lies in the outer belt.
  While most of the mass of C-types is in the middle and outer belt,
  the inner belt and Cybeles contain roughly the same amount of C-type
  material (1\%). P-types are actually relatively evenly distributed 
  throughout the middle, outer, and Cybele regions and the Hildas and
  Trojans only account for a small mass ($<$15\%) although this
  number is biased because discoveries among Hildas and Trojans
  are incomplete. 

  \indent \add{Tables~\ref{tab: masstot},~\ref{tab: masszone}, and~\ref{tab: massclass}
    provide the total amount of material for each asteroid class
    present in the inner solar system and how they are distributed.
    The composition of each body was set at an early stage of the solar
    system formation when the asteroids accreted. The subsequent dynamical history of
    the solar system may have shifted their positions and greatly reduced their
    numbers \citep[e.g.,][]{2005-Nature-435-Gomes,2005-Nature-435-Morbidelli,2005-Nature-435-Tsiganis,2009-Nature-457-Minton}.
    The masses listed in Table~\ref{tab: masstot} can be directly
    compared to the output of numerical simulations using assumptions on the
    original formation location of  each class.}

\section{Conclusion \label{sec: conc}}

  In this work we present the bias-corrected taxonomic distribution
  of asteroids down to 5\,km.
  \add{We present a method to connect the broad-band photometry of the Sloan
    Digital Sky Survey to previous asteroid taxonomies, based on spectra with
    high spectral resolution and similar wavelength range.
    Such a method could be applied to other multi-filter surveys.}
  We \add{then present a bias-correction} method relevant
  to large datasets whereby we select the least-biased subset
  to account for regions and sizes adequately sampled 
  \add{by the SDSS survey}
  and include a correction for discovery incompleteness of the 
  MPC at the smallest sizes.
  The color-based taxonomy and bias-correction is used to study the
  distribution of material
  in the asteroid belt for the first time according to a variety of new parameters 
  (surface area, volume, and mass) rather than number to more
  accurately represent the total material. 
  \add{These quantities add a new perspective that was previously unachievable 
    by studying solely the distribution by number. 
    For instance, the E-types, are often described as unique to the Hungarias
    yet 90\% of E-type mass is in the main belt. Additionally, the primary residence 
    of S-types is typically thought to be the inner belt, yet we find half the mass of S-types 
    is in the middle belt.} 

  We confirm many trends seen in previous works \add{with S- and V-type 
    asteroids accounting for most of the inner belt, and C-, P-, and D-type asteroids
    dominating the outer belt to the Trojans. We find this 
    view of the compositional distribution of the largest bodies determined in the early
    1980s is still robust.}
  We confirm the presence of S-types in the outer parts of the
  main belt as seen by \citet{2003-Icarus-162-Mothe-Diniz} as well as the
  scarcity of S-type among Hildas and Trojans noted by many authors.
  \add{We find evidence for numerous D-types in the inner belt that were previously expected to be
  nonexistent in that region, and in possible contradiction with
  the dynamical models of implementation of trans-Neptunian objects in the
  outer belt during planetary migration}. 
  
  The main belt's most massive classes are C, B, P, V and S in 
    decreasing order. Excluding the four most massive asteroids, (1) Ceres, (2) Pallas, (4) Vesta and
    (10) Hygiea that heavily skew the values, 
    primitive material (C-, P-types) account for more than half main-belt and
    Trojan asteroids, most of the remaining mass being in the S-types. All the
    other classes are minor contributors to the material between Mars and Jupiter.

\section*{Acknowledgments}
  We thank Rick Binzel, Tom Burbine, and Andy Rivkin for useful
  discussions and clarifications. We thank two anonymous referees for
  helpful comments. We acknowledge support from the Faculty of
  the European Space Astronomy Centre (ESAC) for F.D.'s
  visit.  This material is based upon work supported by the
  National Science Foundation under Grant 0907766 and by the National Aeronautics and
  Space Administration under Grant No. NNX12AL26G. Any opinions, findings, and
  conclusions or recommendations expressed in this material are those of the
  authors and do not necessarily reflect the views of the National Science
  Foundation or the National Aeronautics and Space Administration
  
  Funding for the SDSS and SDSS-II has been provided by the Alfred P. Sloan
  Foundation, the Participating Institutions, the National Science
  Foundation, the U.S. Department of Energy, the National Aeronautics and
  Space Administration, the Japanese Monbukagakusho, the Max Planck Society,
  and the Higher Education Funding Council for England. The SDSS Web Site is
  http://www.sdss.org/. 
  
  This publication makes use of data products from the Wide-field Infrared
  Survey Explorer, which is a joint project of the University of California,
  Los Angeles, and the Jet Propulsion Laboratory/California Institute of
  Technology, funded by the National Aeronautics and Space
  Administration.



\end{document}